\documentclass{siamart220329}
\usepackage{calrsfs}
\usepackage{mathrsfs}
\usepackage{enumerate}

\usepackage{graphicx}
\usepackage{multirow}
\usepackage{amsmath,amssymb,amsfonts}
\usepackage{mathrsfs}
\usepackage[title]{appendix}
\usepackage{xcolor}
\usepackage{textcomp}
\usepackage{manyfoot}
\usepackage{booktabs}
\usepackage{algorithm}
\usepackage{algorithmicx}
\usepackage{algpseudocode}
\usepackage{listings}
\usepackage{upgreek}

\newtheorem{remark}[theorem]{Remark}

\usepackage{hyperref}

\hypersetup{
  colorlinks,
  citecolor=blue,
  linkcolor=blue,
  urlcolor=blue}

\newcommand{\su}{\mathtt{u}}
\renewcommand{\Re}[1]{\textrm{~Re~}#1}
\renewcommand{\Im}[1]{\textrm{~Im~}#1}
\DeclareMathOperator{\Ld}{\mathtt{L}}
\DeclareMathOperator{\sw}{0}
\newcommand{\tldL}{\tilde{L}}
\newcommand{\tldR}{\tilde{R}}
\newcommand{\tldT}{\tilde{T}}
\newcommand{\gammap}{\upsilon}
\newcommand{\Dsf}{\mathcal{D}}

\title{Surface waves in randomly perturbed discrete 
models
}
\author{Josselin Garnier\thanks{Centre de Math\'{e}matiques Appliqu\'{e}es, Ecole Polytechnique, Institut Polytechnique de Paris, Palaiseau, 91120, France (\email{josselin.garnier@polytechnique.edu}).}
\and Basant Lal Sharma\thanks{Department of Mechanical Engineering, Indian Institute of Technology Kanpur, Kanpur, 208016, UP, India  (\email{bls@iitk.ac.in}).}
}

\date{\today}

\begin{document}

\maketitle

\begin{abstract}{We study the propagation of surface waves across structured surfaces with random, localized inhomogeneities.
A discrete analogue of Gurtin-Murdoch model is employed and surface elasticity, in contrast to bulk elasticity, is captured by distinct 
point masses and elastic constants for nearest-neighbour interactions parallel to the surface.
Expressions for the surface wave reflectance and transmittance, as well as the radiative loss,
are provided for every localized patch of point mass perturbation on the surface.
As the main result in the article, we provide the statistics of surface wave reflectance and transmittance and the radiative loss for an ensemble of random mass perturbations, independent and identically distributed with mean zero, on the surface.
In the weakly scattering regime, the mean radiative loss is found to be proportional to the size of the perturbed patch, to the variance of the mass perturbations, and to an effective parameter that depends on the continuous spectrum of the unperturbed system.
In the strongly scattering regime, the mean radiative loss is found to depend on another effective parameter that depends on the continuous spectrum, but not on the variance of the mass perturbations.
Numerical simulations are found in quantitative agreement with the theoretical predictions for several illustrative values of the surface structure parameters.}
\end{abstract}

 

\section{Introduction}

Surface waves play an important role at small length scales in physical structures, in addition to their role in traditional engineering systems \cite{Smolyaninov,
Maurer,
Zhang20,
Bakre,
Dixon}.
However, surfaces and coatings are not perfect \cite{Merlitz,Variola} and this leads to several challenges along with their diverse applications at nano-scale, though sometimes defects are intentionally crafted for desired signal transmission.
The high frequency wave scattering, due to steps and serrations on crystal \cite{MartinJ,Schiller}, involves the discreteness of structure and the perturbations influence the dynamical and wave phenomena \cite{Gasteau,Dransfeld,Wang2019}.
Indeed, the physics of surfaces and interfaces has evolved over several decades \cite{Wallis,Sakuma,Steg,Wallis1979} and the nano-scale effects are becoming increasingly relevant in current science such as the case of functional nanomaterials and miniaturized structures \cite{Schaefer,Reetz,Besenbacher} for their elastic, electronic and thermal properties. 
On the other hand, the presence of randomness at small scales has led to an impact on the fundamental physical principles
\cite{Anderson,Chaudhuri}.

Restricting to the classical theories, within the continuum models of surface elasticity, \cite{GurtinMurdoch1975a,Gurtin2} proposed an independent set of constitutive relations on the surface in addition to the ones defined in the bulk; this modeling can be interpreted as an elastic membrane glued on the material surface and is referred as the Gurtin-Murdoch model.
Recently, a discrete Gurtin-Murdoch model was introduced by \cite{Victor_Bls_surf1} wherein a comparison of the surface waves between the continuous Gurtin-Murdoch model and the discrete framework of lattice dynamics is also provided within the kinematic assumption of anti-plane shear deformation (out-of-plane motion). 
The problem of scattering of such surface waves, due to specific surface inhomogeneities, has been tackled in \cite{Victor_Bls_surf2,Blstgsurf,Blssurf} where it is found that the energy flux of surface waves is `leaked' at the defects in the form of bulk waves. 
In more general situations and complicated lattices, indeed, it is quite common to find surface (phonon) bands in the crystal models \cite{Brillouin,Maradudin,Wallis} though the calculations in the lines of \cite{Blstgsurf,Blssurf} are less convenient.

\begin{figure}[htb!]
\centering
\includegraphics[width=.99\textwidth]{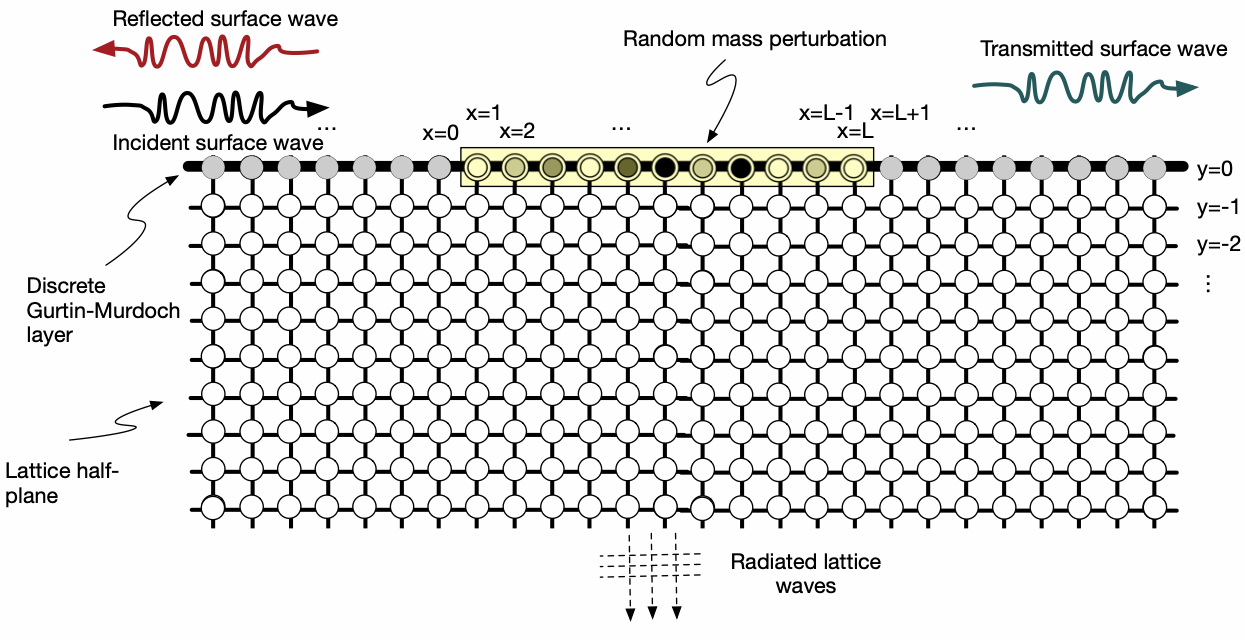}
\caption{Schematic of surface wave transmission in a randomly perturbed Gurtin-Murdoch lattice model: a square lattice half plane endowed with a boundary with structure containing certain random mass defects in a finite region of size $\Ld$ sites.}
\label{Fig1sqhalf}
\end{figure}

Specifically, in this article, a prototype of surface wave propagation in a semi-infinite square lattice half plane representing surface of a crystalline structure is analyzed in the presence of random perturbations of the masses at the boundary.
The specific {lattice} model adopted, based on that introduced in \cite{Victor_Bls_surf1}, is shown schematically in Fig. \ref{Fig1sqhalf}.
The techniques developed in the general problem of discrete scattering on uniform square lattices \cite{sK,sC,sharma2017scattering,BlsMaurya3}, without additional surface structure, do assist in carrying forward the analysis to such structured half planes while our recent work on one dimensional lattice model \cite{JGarnierBls} paved way for the choice of the simplest form of half plane model.
As a preliminary result, the exact solution of the problem of scattering of surface waves due to a patch of size $\Ld$ of perturbed masses is provided using the lattice Green's function for the assumed structure; the latter corresponds to the solution of the discrete Helmholtz equation in the semi-infinite lattice with a point source at the surface \cite{Blssurf}. 
The reflection and transmission coefficients for surface wave propagation, from one side of the patch to another, can then be obtained via the inversion of a $\Ld\times \Ld$ matrix. 
Together with a conservation of energy relation, the expressions of the reflection and transmission coefficients give in turn a formula for the radiative loss, namely the fraction of incident energy flux that is `leaked' at the interface in the form of bulk waves.
The main result concerns the question about the dependence of the physically important effects of randomness in the surface material parameters. 
This is answered for surface wave transmission problem by an application of stochastic and multiscale analysis \cite{book}.
Such an approach has already been successfully applied to wave propagation in randomly perturbed systems such as acoustic, elastic or electromagnetic waves, including the wave propagation in randomly perturbed waveguides \cite{papa07,borcea14,borcea17} and the surface wave propagation in a randomly perturbed continuous half-space \cite{Hoop,Borcea}.
In the case of the randomly perturbed discrete Gurtin-Murdoch model addressed in this paper, it is possible to study the regime where the relative standard deviation of the mass perturbations is small and the size $\Ld$ of the perturbed patch is large so that the cumulative effects of the mass perturbations onto the propagation of an incident surface wave are of order one. 
A system of stochastic differential equations for the reflection and transmission coefficients as functions of the size of the perturbed patch can be obtained, which takes into account all relevant phenomena, including the coupling with radiative and evanescent modes. 
This system gives in turn the expressions of the moments of the reflection and transmission coefficients and the radiative loss. 
An inspection of these expressions reveals that, in the weakly scattering regime, the mean radiative loss is proportional to the size of the randomly perturbed patch, to the variance of the mass perturbations, and to an effective parameter that depends on the continuous spectrum of the unperturbed system.
In the strongly scattering regime, the mean radiative loss depends on another effective parameter, but not on the variance of the mass perturbations.
The numerical calculations suggest that the dependence of transmittance and radiative loss on the incident wave frequency is non-monotone in surface wave band for each instance of perturbation while the mean values are monotone mostly. 
In general, the results on transmittance and radiative loss are highly susceptible to the choice of values of physical parameters associated with the unperturbed surface structure. 

The present article is organized as follows. The discrete Gurtin-Murdoch lattice model incorporating the surface structure is formulated in \S\ref{secGMmodel} along with a short description of the associated surface waves. 
\S\ref{secGMmodel} provides the complete eigenbasis for the difference operator characterizing the unperturbed one dimensional semi-infinite lattice model closely related to the lattice half plane.
\S\ref{secGMmodel} also contains details on the modal expansion of the solution of the unperturbed equation of motion of lattice half plane with surface structure.
The exact solution of scattering problem of surface waves, due to a finite size of patch, of possibly large, mass perturbation on the boundary, appears in \S\ref{secexact}. 
The expressions of reflection and transmission coefficients are provided in \S\ref{secexact} using the Green's function for point source on the boundary of half plane. 
\S\ref{secexact} also contains the energy conservation theorem which provides the statement of distribution of the energy flux of incident surface wave into reflected and transmitted surface waves apart from the waves radiated into the half plane.
The main result of this article appears in \S\ref{secmain} concerning the effects on scattering of surface waves due to randomness in small mass perturbation on a finite sized patch on the boundary.
Proof of the main result is provided in \S\ref{secmainproof} using a multiscale analysis of the the modal expansion of the solution of the perturbed equation of motion, dealing with the role of the evanescent and radiative modes.
After concluding remarks, a list of references and two appendices complete the article.

{\bf Mathematical preliminaries:}
Let ${\mathbb{Z}}$ denote the set of integers, ${{\mathbb{Z}}^2}$ denote ${\mathbb{Z}}\times{\mathbb{Z}},$
and ${{\mathbb{Z}}^-}$ denote the set of negative integers.
The definitions
\begin{subequations}
 \begin{equation}
{\mathbb{H}}:=\{({{\mathtt{x}}}, {{\mathtt{y}}})\in{{\mathbb{Z}}^2}: {{\mathtt{y}}}\le0\},\quad
\label{hlfplane1}
\end{equation}
\begin{equation}
\mathring{\mathbb{H}}:=\{({{\mathtt{x}}}, {{\mathtt{y}}})\in{{\mathbb{Z}}^2}: {{\mathtt{y}}}\in\mathbb{Z}^-\},\quad
\partial{\mathbb{H}}:=\{({{\mathtt{x}}}, {{\mathtt{y}}})\in{{\mathbb{Z}}^2}: {{\mathtt{y}}}=0\},
\label{hlfplane2}
\end{equation}
\end{subequations}
represent the square lattice {half-plane} with boundary, without boundary and the boundary itself, respectively. 
Let ${\mathbb{R}}$ denote the set of real numbers and ${\mathbb{C}}$ denote the set of complex numbers. 
{For} ${z}\in\mathbb{C}$,
$\Re {z}\in{\mathbb{R}}$ denotes the real part,
$\Im {z}\in{\mathbb{R}}$ denotes the imaginary part, 
{$\overline{z}$ denotes the complex conjugate $\Re {z}-i \Im {z}$ of ${z}$,}
$|{z}|$ denotes the modulus, and $\arg {z}$ denotes the principal argument.
The square root function has the usual branch cut in the complex plane running from $-\infty$ to $0$. 
The symbol ${\mathbb{T}}$ denotes the unit circle (as a counterclockwise contour) in the complex plane.
If $f$ is a differentiable, real or complex valued, function then $f'$ denotes the derivative of $f$. 
The decoration $~\check{}~$ is used to represent the modal amplitudes in contrast to $~\hat{}~,~ \widehat{}~$ for the incident wave.
The subscript $0$ typically accompanies entities related to the surface wave while the subscript $s$ appears with surface structure parameters. The usage of ${{\mathtt{x}}}, {{\mathtt{y}}}$ is restricted to emphasize the lattice coordinates; in particular, ${\mathtt{x}}, {\mathtt{y}}\in\mathbb{Z}$.
The symbol ${\mathfrak{z}}$ is reserved for a complex variable.

\section{Lattice model and completeness of eigenfunctions}
\label{secGMmodel}

\subsection{Discrete Gurtin-Murdoch lattice model}
Consider the unperturbed lattice model, ignoring the yellow patch in Fig. \ref{Fig1sqhalf} which provides a schematic representation of the lattice model with surface structure and point mass perturbation. 
Within the paradigm of the out-of-plane displacement field of a lattice structure, 
the equation of motion on ${\mathring{\mathbb{H}}}$, the portion of lattice half-plane away from its boundary, 
is 
\begin{subequations}
 \begin{equation}\begin{split}
{\triangle}{\su}_{{{\mathtt{x}}}, {{\mathtt{y}}}}+{\omega}^2{\su}_{{{\mathtt{x}}}, {{\mathtt{y}}}}=0,\quad {({{\mathtt{x}}},{{\mathtt{y}}})\in{\mathring{\mathbb{H}}}},\\
\text{where }
{\triangle}{\su}_{{{\mathtt{x}}}, {{\mathtt{y}}}}\equiv{\su}_{{{\mathtt{x}}}+1, {{\mathtt{y}}}}+{\su}_{{{\mathtt{x}}}-1, {{\mathtt{y}}}}+{\su}_{{{\mathtt{x}}}, {{\mathtt{y}}}+1}+{\su}_{{{\mathtt{x}}}, {{\mathtt{y}}}-1}-4{\su}_{{{\mathtt{x}}}, {{\mathtt{y}}}}.
\label{eqbulk}
\end{split}\end{equation}
The equation of motion of the particles on the lattice half-plane boundary $\partial{\mathbb{H}}$, as the counterpart of \eqref{eqbulk}, is
\begin{equation}\begin{split}
\alpha_{{s}}({\su}_{{{\mathtt{x}}}+1, {{\mathtt{y}}}}+{\su}_{{{\mathtt{x}}}-1, {{\mathtt{y}}}}-2{\su}_{{{\mathtt{x}}}, {{\mathtt{y}}}})+{\su}_{{{\mathtt{x}}}, {{\mathtt{y}}}-1}-{\su}_{{{\mathtt{x}}}, {{\mathtt{y}}}}+m_{{s}}{\omega}^2{\su}_{{{\mathtt{x}}}, {{\mathtt{y}}}}=0, \quad {\mathtt{x}}\in\mathbb{Z}, {\mathtt{y}}=0,
\label{eqsurf}
\end{split}\end{equation}
\end{subequations}
where 
$m_{{s}}>0$ represent the possibly distinct mass of particles at $\partial{\mathbb{H}}$ and $\alpha_{{s}}>0$ denotes the elastic constants for nearest-neighbour interactions parallel to the surface.
Eqs. \eqref{eqbulk}, \eqref{eqsurf} can be postulated following the physical scaling described in Section 1 of \cite{Blssurf}.

The above equations, \eqref{eqbulk} in $\mathring{\mathbb{H}}$ and \eqref{eqsurf} in $\partial{\mathbb{H}}$, allow a surface wave band for the values of $\alpha_{{s}}$ and $m_{{s}}$ such that $\alpha_{{s}}<m_{{s}}$ \cite{Victor_Bls_surf1}; see \cite{Blssurf} for an elementary justification of this condition and, in particular, note that the case $m_{{s}}=\alpha_{{s}}=1$ does not permit such surface wave band.
Indeed, it is easy to see, after the substitution of
\begin{equation}\begin{split}
{\su}_{{{\mathtt{x}}},{{\mathtt{y}}}}=e^{ i{{\mathit k}}_{\sw} {{\mathtt{x}}}+{\eta}({{\mathit k}}_{\sw}) {{\mathtt{y}}}},\quad {\mathit k}_{\sw}\in(-\pi,\pi)\setminus\{0\},
\label{surfwavemode}
\end{split}\end{equation}
in \eqref{eqbulk},\eqref{eqsurf}, that the common description of surface waves is satisfied,
provided
${{\omega}}\equiv{{\omega}}({{\mathit k}}_{\sw})$
and ${\eta}\equiv{\eta}({{\mathit k}}_{\sw})>0$ are implicitly obtained from the two coupled equations:
\begin{align}\label{funcomegaeq}
{{\omega}}^2&=4\sin^2({{{\mathit k}}}_{\sw}/{2})+2-2\cosh{\eta},\quad
{m}_{{s}}{{\omega}}^2=4{\alpha}_{{s}} \sin^2({{{\mathit k}}_{\sw}}/{2})+1-e^{-{\eta}}.
\end{align}
Thus, ${\su}$ in \eqref{surfwavemode} represents a {\em surface wave mode} of dimensionless wavenumber ${{\mathit k}}_{\sw}\in(-\pi,\pi)\setminus\{0\}$ that
decays exponentially into $\mathbb{H}$ with an attenuation coefficient ${{\eta}({{\mathit k}}_{\sw})}$. 
Note that 
\begin{align}
{\omega}({{\mathit k}}_{\sw})\in(0, {\omega}_{\max})=:\mathcal{P}_{{{s}}},\quad {{\mathit k}}_{\sw}\in(-\pi,\pi)\setminus\{0\},
\label{PBdef}
\end{align}
where ${\omega}_{\max}$ is
given by
\begin{align}
&\omega_{\max} =\sqrt{(-b_s+\sqrt{b_s^2-4a_sc_s})/(2a_s)},\label{defwmax}
\\
\text{with }a_s&=m_s(m_s-1),\quad 
b_s=-8\alpha_sm_s+4\alpha_s+4m_s+1,\nonumber \\
&c_s=16\alpha_s^2-16\alpha_s-4.\nonumber
\end{align}
For each given $\omega\in\mathcal{P}_{{{s}}}$, there exist two surface wave modes of the form \eqref{surfwavemode}, which can be classified as a left-going and a right-going wave based on the sign of the propagation of energy flux.
The group velocity of {the surface wave} \eqref{surfwavemode} with wave number ${{\mathit k}}_{\sw}\in(-\pi,\pi)\setminus\{0\}$ is \cite{Brillouin}
\begin{align}
{\mathit v}_{{{s}}}({{{\mathit k}}_{\sw}}):=
\frac{d}{d\xi}{{\omega}}(\xi)|_{\xi={{\mathit k}}_{\sw}}
=\frac{\sin{{\mathit k}}_{\sw}}{{\omega}({{{\mathit k}}_{\sw}})}\frac{{\alpha}_{{{s}}}+(e^{2{{{\eta}({{{\mathit k}}_{\sw}})}}}-1)^{-1}}{{m}_{{{s}}}+(e^{2{{{\eta}({{{\mathit k}}_{\sw}})}}}-1)^{-1}}.
\label{surfgpvel}
\end{align}
The surface wave \eqref{surfwavemode} carries energy flux towards ${{\mathtt{x}}}\to+\infty$ (resp. $-\infty$) when ${{{\mathit k}}_{\sw}}\in(0,\pi)$ (resp. $-{{\mathit k}}_{\sw}\in(0, \pi)$) as ${\mathit v}_{{{s}}}({{{\mathit k}}_{\sw}})>0$ (resp. ${\mathit v}_{{{s}}}({{{\mathit k}}_{\sw}})<0$).

Before proceeding to the main results the randomly perturbed discrete Gurtin-Murdoch model with localized mass perturbations on sites in $\partial{\mathbb{H}}$, certain results regarding the spectral properties of the difference operator involved in \eqref{eqbulk},\eqref{eqsurf} on $\mathbb{H}$ are needed.
In retrospect, it is convenient to study a reduced one dimensional lattice model on $\mathbb{Z}^-\cup\{0\}$ at first.

\subsection{Completeness of eigenfunctions for a (reduced) one dimensional semi-infinite lattice model}
\label{seccomplete}
Suppose that ${\omega}$ belongs to the regime when a surface wave mode \eqref{surfwavemode} exists on $\mathbb{H}$ and \eqref{eqbulk},\eqref{eqsurf} are satisfied, i.e. ${\omega}\in(0, {\omega}_{\max})\subset(0,2)$ where ${\omega}_{\max}$ is given by \eqref{defwmax}.
Following Appendix A of \cite{Hoop}, the objective is to show that the solution of the unperturbed system can be expanded on a complete system of eigenfunctions, which represent both the propagative modes, that is surface mode and the radiative modes, and the evanescent modes.

We first introduce the unnormalized functions $\psi_\gamma$ which appear in the expressions of the eigenfunctions. 
\subsubsection{Unnormalized functions}
\label{secunnorm}
For any $\gamma \in \mathbb{R}$, we denote by $\psi_\gamma:\mathbb{Z}^-\cup\{0\}\to\mathbb{C}$ the unique solution of the second-order difference equation 
\begin{subequations}
 \begin{equation}
\begin{split}
(\Delta_1 +{\omega}^2)\psi_\gamma({\mathtt{y}})=\gamma\psi_\gamma({\mathtt{y}}),\quad {\mathtt{y}}\in\mathbb{Z}^-,
\label{eqnhalf}
\end{split}
\end{equation}
\begin{equation}
\begin{split}
\psi_\gamma({\mathtt{y}}-1)-\psi_\gamma({\mathtt{y}}) +{m}_{{s}}{\omega}^2\psi_\gamma({\mathtt{y}})={\alpha}_{{s}}\gamma\psi_\gamma({\mathtt{y}}),\quad {\mathtt{y}}=0,
\label{bc1}
\end{split}
\end{equation}
\begin{equation}\begin{split}
\psi_\gamma({\mathtt{y}})=1,\quad {\mathtt{y}}=0,
\label{bc2}
\end{split}\end{equation}
where
$\Delta_1$ is the one dimensional discrete Laplacian
\begin{equation}\begin{split}
{\Delta_1}\psi({\mathtt{y}})\equiv\psi({\mathtt{y}}+1)+\psi({\mathtt{y}}-1)-2\psi({\mathtt{y}}), \quad {\mathtt{y}}\in\mathbb{Z}, \psi:\mathbb{Z}\to\mathbb{C}.
\end{split}\end{equation}
\end{subequations}

The function $\psi_\gamma$ has the following form:\\
If $\gamma \neq {\omega}^2$, then there exists a unique $\beta\neq 0$ such that $ \Re\beta\ge0$, $\Im\beta\ge0$, and $e^{\beta}+e^{-\beta}-2+{\omega}^2=\gamma$. Then
$\psi_\gamma=A_\gamma e^{\beta {\mathtt{y}}}+(1-A_\gamma )e^{-\beta {\mathtt{y}}}$, where
$A_\gamma = ({-\alpha_{{s}} \gamma -1 +m_{{s}}{\omega}^2+e^{\beta}})/({e^{\beta}-e^{-\beta}})$
(so that the two boundary conditions \eqref{bc1}, \eqref{bc2} hold).\\
If $\gamma={\omega}^2$, then $\psi_\gamma=1+A_\gamma {\mathtt{y}}$, where
$A_\gamma= (m_{{s}}-\alpha_{{s}}){\omega}^2$
(so that \eqref{bc1}, \eqref{bc2} hold).

\begin{enumerate}[(i)]
\item 
If $\gamma\in({\omega}^2-4,{\omega}^2)$,
we have 
$\beta=\pm i \zeta, \zeta\in(-\pi/2, \pi/2)$ with $2\cos \zeta-2+{\omega}^2=\gamma$.
For ${\mathtt{y}}\leq 0$, 
$\psi_\gamma=A_\gamma e^{i\zeta {\mathtt{y}}} +\overline{A_\gamma} e^{- i\zeta {\mathtt{y}}} $, 
with $A_\gamma=\frac{1}{2}+\frac{i}{2\sin \zeta} (-\cos \zeta +\alpha_{{s}} \gamma +1-m_{{s}} {\omega}^2)$,
which yields
\begin{equation}\begin{split}
\psi_\gamma({\mathtt{y}})=\frac{1}{\sin\zeta}\big( ({m}_{{s}} {\omega}^2-{\alpha}_{{s}} \gamma-1 )\sin\zeta {\mathtt{y}}+\sin\zeta ({\mathtt{y}}+1)\big).
\label{defzetagm}
\end{split}\end{equation}
Thus, $\psi_\gamma$ is bounded on $\mathbb{Z}^-$.
 
\item
If $\gamma<{\omega}^2-4$ or $\gamma>{\omega}^2$, then $\psi_\gamma$ is exponentially growing on $\mathbb{Z}^-$ as ${\mathtt{y}}\to-\infty$ because $A_\gamma\neq 1$ except for one value $\gamma_0 >{\omega}^2$ (assuming ${\alpha}_{{s}}<m_{{s}}$) which is such that $A_{\gamma_0}=1$.
This value $\gamma_0 $ yields an eigenfunction $\psi_{\gamma_0}({\mathtt{y}})=e^{\beta_0 {\mathtt{y}}}$ where $\beta_0>0$ is such that 
\begin{equation}\begin{split}
{\omega}^2=2+\gamma_0-2\cosh\beta_0,\qquad
{m}_{{s}} {\omega}^2={\alpha}_{{s}} \gamma_0 +1-e^{-\beta_0}.
\label{localeigenfn}
\end{split}\end{equation}
Thus, $\psi_{\gamma_0}$ is square 
summable on $\mathbb{Z}^-$.
\end{enumerate}

\subsubsection{Orthonormal eigenfunctions}
\label{secnorm}
We consider the weighted $l^2$ space of typical $u:\mathbb{Z}^-\cup\{0\}\to\mathbb{C}$ with the norm defined by
\begin{equation}
\| u \|_2^2:= \alpha_{{s}} |u_0|^2 + \sum_{{\mathtt{y}}\in\mathbb{Z}^-} |u_{{\mathtt{y}}}|^2,
\label{l2norm}
\end{equation}
and the associated inner product defined by
\begin{equation}
\langle u,v\rangle := \alpha_{{s}} u_0 \overline{v_0} +\sum_{{\mathtt{y}}\in\mathbb{Z}^-} u_{{\mathtt{y}}} \overline{v_{{\mathtt{y}}}}.
\label{l2inner}
\end{equation}
In this context, recall that $\alpha_{{s}}>0$ so that above definitions are non-degenerate.
Inspired by the equations \eqref{eqnhalf}, \eqref{bc1}, \eqref{bc2},
consider the definition:
\begin{equation}
(\mathcal {L} u)_{{\mathtt{y}}} := \left\{
\begin{array}{ll}
\Delta_1 u_{{\mathtt{y}}} +{\omega}^2 u_{{\mathtt{y}}} &\mbox{ if } {\mathtt{y}}\in\mathbb{Z}^-,\\
\frac{1}{\alpha_{{s}}} (u_{-1}-u_0+m_{{s}}{\omega}^2 u_0) &\mbox{ if } {\mathtt{y}}=0 .
\end{array}
\right.
\label{defLoper}
\end{equation}
By a direct calculation, it is found that the operator $\mathcal {L}$ defined by \eqref{defLoper}
is self-adjoint in $l^2$, i.e.
\[
\langle u,\mathcal {L}v\rangle=\langle\mathcal {L}u,v\rangle,
\]
for all $u,v\in l^2$ with weighted inner product defined by \eqref{l2inner}.

A sketch of the (standard) proof of spectral properties of $\mathcal {L}$ is as follows.
Let us fix some positive integer $D$. If one considers the domain $[-D,0]\cap\mathbb{Z}$ with Dirichlet boundary condition at ${\mathtt{y}}=-D$ instead of $(-\infty,0]\cap\mathbb{Z}$, then the operator $\mathcal {L}$ is self-adjoint and compact (equivalent to a $D\times D$ matrix in this case). 
By the spectral theorem for $D\times D$ symmetric matrices there exists an orthonormal basis consisting of eigenfunctions of $\mathcal {L}$. 
The eigenvalues are simple $\gamma_{j,D}$ and the eigenfunctions $\phi_{\gamma_{j,D}}$ (proportional to $\psi_{\gamma_{j,D}}$) are orthogonal. 
One can then proceed as in \cite[Appendix A]{Hoop}. By taking the limit $D\to+\infty$ one gets that the operator $\mathcal {L}$ on $l^2$ has a discrete spectrum made of one isolated eigenvalue $\gamma_0$ and a continuous spectrum on $({\omega}^2-4,{\omega}^2)$. 

\begin{proposition}
Any function $u\in l^2$ (with weighted inner product \eqref{l2inner}) can be expanded as \begin{equation}
u_{{\mathtt{y}}} = \check{u}_{\gamma_0} \phi_{\gamma_0}({\mathtt{y}}) +\int_{{\omega}^2-4}^{{\omega}^2} \check{u}_\gamma \phi_\gamma({\mathtt{y}}) d\gamma,
\end{equation}
with 
the modal amplitudes given by
\begin{subequations}
 \begin{align}
&\check{u}_{\gamma_0} = \langle u,\phi_{\gamma_0}\rangle,\label{ucheck0}\\
&\check{u}_\gamma = \langle u,\phi_{\gamma}\rangle,\quad
\gamma \in ({\omega}^2-4,{\omega}^2),\label{ucheck}
\end{align}
and eigenmodes given by
\begin{align}
&\phi_{\gamma_0} ({\mathtt{y}}) = \sqrt{\rho( \gamma_0)} \psi_{\gamma_0} ({\mathtt{y}}),\\
&\phi_{\gamma} ({\mathtt{y}}) = \sqrt{\rho(\gamma)} \psi_{\gamma} ({\mathtt{y}}),\quad \gamma \in ({\omega}^2-4,{\omega}^2), 
\end{align}
where
\begin{align}
&\rho( \gamma_0) = \frac{1}{ \alpha_{{s}} +(e^{2\beta_0}-1)^{-1}},\label{rhogm0}\\
&\rho(\gamma) = \frac{2}{\pi} \frac{\sqrt{({\omega}^2-\gamma)(\gamma-{\omega}^2+4)}^{-1}}{1+ \frac{(\gamma(2\alpha_{{s}}-1)+{\omega}^2(1-2m_{{s}}))^2}{({\omega}^2-\gamma)(\gamma-{\omega}^2+4)}},\quad \gamma \in ({\omega}^2-4,{\omega}^2).\label{rhogm}
\end{align}
\label{ucheckdefns}
\end{subequations}
We, also, have the Parseval relation
\begin{equation}
\|u\|_2^2 = |\check{u}_{\gamma_0}|^2 +\int_{{\omega}^2-4}^{{\omega}^2} |\check{u}_\gamma|^2 d\gamma 
\end{equation}
for all $u\in l^2$.
\label{propcomp}
\end{proposition}
Proposition \ref{propcomp} shows that, after suitable normalization of $\psi_\gamma$ obtained in Section \ref{secunnorm},
a function in $l^2$ can be expanded on the complete set of eigenfunctions $\psi_\gamma$,
$\gamma \in ({\omega}^2-4,{\omega}^2) \cup \{\gamma_0\}$.
An alternative way to arrive at the normalization of $\psi_\gamma$ is provided in Appendix \ref{app:secGreen}.

\subsection{Modal expansion of the solution of the unperturbed equation of motion}
In view of Proposition \ref{propcomp},
the solution ${\su}:\mathbb{H}\to\mathbb{C}$,
which satisfies the equation of motion \eqref{eqbulk} with the surface condition \eqref{eqsurf} on $\partial{\mathbb{H}}$ for all ${\mathtt{x}}\in\mathbb{Z}$, can be expanded as the superposition of orthonormal modes obtained in Section \ref{secnorm}:
\begin{align} 
&{\su}_{{\mathtt{x}},{\mathtt{y}}}=\check{\su}_{\gamma_0} ({\mathtt{x}})\phi_{\gamma_0}({\mathtt{y}})+\int_{{\omega}^2-4}^{{\omega}^2}\check{\su}_\gamma({\mathtt{x}}) \phi_\gamma({\mathtt{y}}) d\gamma,
\quad ({{\mathtt{x}}}, {{\mathtt{y}}})\in{\mathbb{H}},
\label{eq:expandsol}\\
&\textrm{with }
\check{\su}_{\gamma_0}({\mathtt{x}}) = \langle {\su}_{{\mathtt{x}},\cdot},\phi_{\gamma_0}\rangle,\\
&\hspace*{0.36in} \check{\su}_\gamma({\mathtt{x}}) = \langle {\su}_{{\mathtt{x}},\cdot},\phi_{\gamma}\rangle,\quad \gamma \in ({\omega}^2-4,{\omega}^2),
\end{align}
using the weighted inner product defined by \eqref{l2inner}. Note that $\{{\su}_{{\mathtt{x}},{\mathtt{y}}}\}_{{\mathtt{y}}\in\mathbb{Z}^-}$ belongs to $l^2$ for each ${\mathtt{x}}\in\mathbb{Z}.$

The modal amplitudes in \eqref{eq:expandsol} satisfy the uncoupled difference equations
\begin{equation}
\check{\su}_\gamma({\mathtt{x}}+1)+\check{\su}_\gamma({\mathtt{x}}-1)+(\gamma-2) \check{\su}_\gamma({\mathtt{x}}) =0,\quad {\mathtt{x}}\in\mathbb{Z}.
\end{equation}
\begin{enumerate}[(i)]
\item 
If $\gamma \in (0,4)$ then $\check{\su}_\gamma({\mathtt{x}})$ has the form
\[
\check{\su}_\gamma({\mathtt{x}}) = a e^{ik(\gamma){\mathtt{x}}} +b e^{-ik(\gamma){\mathtt{x}}},\quad {\mathtt{x}}\in\mathbb{Z},
\]
with
\begin{equation}
k(\gamma)={\rm arccos}\big(1-\frac{\gamma}{2}\big) \in (0,\pi), \quad \gamma \in (0,4),
\label{defkgm}
\end{equation}
which shows that it is a propagative mode (the mode $ e^{ik{\mathtt{x}}} $ is right-going on $\partial{\mathbb{H}}$, $ e^{-ik{\mathtt{x}}} $ is left-going on $\partial{\mathbb{H}}$).\\
\item 
If $\gamma <0 $ then $\check{u}_\gamma({\mathtt{x}})$ has the form
\[
\check{\su}_\gamma({\mathtt{x}}) = a e^{k(\gamma){\mathtt{x}}} +b e^{-k(\gamma){\mathtt{x}}},\quad {\mathtt{x}}\in\mathbb{Z}, 
\]
with
\begin{equation}
k(\gamma) ={\rm argcosh}\big(1-\frac{\gamma}{2}\big) = \ln\Big( 1-\frac{\gamma}{2}+ \frac{1}{2} \sqrt{ \gamma(\gamma -4)}\Big) >0, \quad \gamma <0,
\end{equation}
which shows that it is an evanescent mode.
\end{enumerate}

To summarize, the discrete eigenvalue $\gamma_0>{\omega}^2$ corresponds to a propagative mode, addressed as a surface wave mode \eqref{surfwavemode} in Section \ref{secGMmodel}, the continuous spectrum has a propagative (radiative) part for $\gamma \in (0,{\omega}^2)$ and an evanescent part for $\gamma \in ({\omega}^2-4,0)$.

\begin{remark}
For the discrete eigenvalue $\gamma_0$, the attenuation coefficient $\beta_0$ in \eqref{localeigenfn} corresponds to ${{{\eta}({{{\mathit k}}_{\sw}})}}$, the wave number $k(\gamma_0)$ in \eqref{defkgm} corresponds to ${{{\mathit k}}_{\sw}}$ in the surface wave mode \eqref{surfwavemode}.
For the continuous spectrum with a propagative (radiative) part $\gamma \in (0,{\omega}^2)$, the wave number $\zeta\equiv\zeta(\gamma)$ in \eqref{defzetagm} and $k\equiv k(\gamma)$ in \eqref{defkgm} correspond to wave vector $(k,\zeta)\in\mathbb{R}^2$ of bulk lattice waves which have the form $e^{ik{\mathtt{x}}+i\zeta{\mathtt{y}}}$ \cite{Brillouin,sK,Blssurf} and satisfies the bulk dispersion relation ${\omega}^2=4\sin^2 \frac{1}{2}k+4\sin^2 \frac{1}{2}\zeta$.
\label{remonedim}
\end{remark}

\section{Exact solution on \texorpdfstring{$\mathbb{H}$}{} for arbitrary localized mass perturbation on \texorpdfstring{$\partial\mathbb{H}$}{}}
\label{secexact}
Suppose $\Ld$ is a given positive integer representing the size of patch on $\partial\mathbb{H}$ containing the mass defect.
Let ${{\upmu}}_{{\mathtt{x}}}$ describe the relative perturbation in mass at the lattice site $({\mathtt{x}}, 0)\in\partial\mathbb{H}.$
For ${\mathtt{x}}\in\{1, 2, \dotsc, \Ld-1, \Ld\}$ as there is a mass defect ${{\upmu}}_{{\mathtt{x}}}m_{{s}}$, the mass of particle is $m_{{s}}(1+{{\upmu}}_{{\mathtt{x}}})$ in place of $m_{{s}}$ in \eqref{eqsurf}. 
Note that the case of arbitrarily large perturbation in point masses associated with surface structure is permitted as long as the mass of individual particles remains positive.

We use ${{\mathit{u}}}_{{\mathtt{x}},{\mathtt{y}}}$ to represent the scattered field in response to an incident surface wave mode $\widehat{\su}_{{\mathtt{x}},{\mathtt{y}}}$ for all $({\mathtt{x}},{\mathtt{y}})\in\mathbb{H}$; naturally,
the total field on $\mathbb{H}$ is given by $\su_{{\mathtt{x}},{\mathtt{y}}}={{\mathit{u}}}_{{\mathtt{x}},{\mathtt{y}}}+\widehat{\su}_{{\mathtt{x}},{\mathtt{y}}}$.

We consider a propagative (surface) wave mode incident from the left side of the surface patch on $\partial\mathbb{H}$ between ${\mathtt{x}}=1$ and ${\mathtt{x}}=\Ld$ (see yellow patch and wave schematic in Fig. \ref{Fig1sqhalf}):
\begin{equation}\begin{split}
\label{incwave0}
\widehat{\su}_{{{\mathtt{x}}},{{\mathtt{y}}}}={{\hat{\rm a}}}~e^{{ i{{\mathit k}}_{\sw} {{\mathtt{x}}}}+{{\eta}({{\mathit k}}_{\sw}) {{\mathtt{y}}}}},\quad {\mathit k}_{\sw}\in(0,\pi),
\end{split}\end{equation}
where ${\hat{\rm a}}\in\mathbb{C}$ is the amplitude; we assume that ${\hat{\rm a}}\ne0$.

We consider the equation of motion \eqref{eqbulk} with the surface condition \eqref{eqsurf} on $\partial\mathbb{H}$ satisfied by the total field ${{\mathit{u}}}_{{{\mathtt{x}}}, {{\mathtt{y}}}}+\widehat{\su}_{{{\mathtt{x}}}, {{\mathtt{y}}}}$ for all ${\mathtt{x}} \leq 0$ and ${\mathtt{x}} \geq \Ld+1$ and the surface condition with perturbed masses, i.e.
$\alpha_{{s}}({{\mathit{u}}}_{{{\mathtt{x}}}+1, {{\mathtt{y}}}}+{{\mathit{u}}}_{{{\mathtt{x}}}-1, {{\mathtt{y}}}}-2{{\mathit{u}}}_{{{\mathtt{x}}}, {{\mathtt{y}}}})+{{\mathit{u}}}_{{{\mathtt{x}}}, {{\mathtt{y}}}-1}-{{\mathit{u}}}_{{{\mathtt{x}}}, {{\mathtt{y}}}}+m_{{s}}(1+{{\upmu}}_{{\mathtt{x}}}) {\omega}^2{{\mathit{u}}}_{{{\mathtt{x}}}, {{\mathtt{y}}}}
+\alpha_{{s}}(\widehat{\su}_{{{\mathtt{x}}}+1, {{\mathtt{y}}}}+\widehat{\su}_{{{\mathtt{x}}}-1, {{\mathtt{y}}}}-2\widehat{\su}_{{{\mathtt{x}}}, {{\mathtt{y}}}})+\widehat{\su}_{{{\mathtt{x}}}, {{\mathtt{y}}}-1}-\widehat{\su}_{{{\mathtt{x}}}, {{\mathtt{y}}}}+m_{{s}}(1+{{\upmu}}_{{\mathtt{x}}}) {\omega}^2\widehat{\su}_{{{\mathtt{x}}}, {{\mathtt{y}}}}
=0$,
is satisfied for ${\mathtt{x}}\in[1,\Ld]\cap\mathbb{Z}$ and ${\mathtt{y}}=0$. 
As the equation of motion \eqref{eqbulk} on $\mathring{\mathbb{H}}$ with the surface condition \eqref{eqsurf} on $\partial\mathbb{H}$ is satisfied by the incident wave $\widehat{\su}$, 
therefore, we find that the scattered wave field ${{\mathit{u}}}_{{{\mathtt{x}}}, {{\mathtt{y}}}}$ needs to satisfy
\eqref{eqbulk} on $\mathring{\mathbb{H}}$ and
at ${\mathtt{y}}=0$,
\begin{equation}\begin{split}
\alpha_{{s}}({{\mathit{u}}}_{{{\mathtt{x}}}+1, {{\mathtt{y}}}}+{{\mathit{u}}}_{{{\mathtt{x}}}-1, {{\mathtt{y}}}}-2{{\mathit{u}}}_{{{\mathtt{x}}}, {{\mathtt{y}}}})+{{\mathit{u}}}_{{{\mathtt{x}}}, {{\mathtt{y}}}-1}-{{\mathit{u}}}_{{{\mathtt{x}}}, {{\mathtt{y}}}}+m_{{s}}{\omega}^2{{\mathit{u}}}_{{{\mathtt{x}}}, {{\mathtt{y}}}}=\sum_{j=1}^{\Ld}\delta_{{\mathtt{x}}, j} f_j,\quad \forall {\mathtt{x}}\in\mathbb{Z},
\label{eqsurf1}
\end{split}\end{equation}
where
\begin{equation}\begin{split}
f_j:=-m_{{s}}{\omega}^2 {{\upmu}}_j({{\mathit{u}}}_{{j}, {0}}+{\widehat{\su}}_{{j}, {0}}),\quad j=1, \dotsc, \Ld.
\label{deffj}
\end{split}\end{equation}

By inspection of \eqref{eqbulk} and \eqref{eqsurf1}, it is clear that the solution can be expressed in terms of a Green's function.
In particular, we consider the Green's function $\mathcal{G}_{{\mathtt{x}},{\mathtt{y}}}$ for 
all $({{\mathtt{x}}}, {{\mathtt{y}}})\in{\mathbb{H}}$
that satisfies \eqref{eqbulk}, \eqref{eqsurf} with a point source 
$\delta_{{\mathtt{x}}, 0}\delta_{{\mathtt{y}}, 0}$.
The Kronecker delta is defined by
\begin{equation}\begin{split}
\delta_{{{\mathtt{x}}}, {n}}:=0, \textrm{ if }{{\mathtt{x}}}\ne n\quad\textrm{ and }\quad \delta_{{{\mathtt{x}}}, {n}}:=1, \textrm{ if }{{\mathtt{x}}}=n, \quad{{\mathtt{x}}}, n\in{\mathbb{Z}}.
\label{defdelta}
\end{split}\end{equation}
The Green's function $\mathcal{G}$ satisfies the following equations:
\begin{subequations}
\begin{equation}\begin{split}\label{eqbulkG}
(\mathcal{G}_{{{\mathtt{x}}}+1,{{\mathtt{y}}}}+\mathcal{G}_{{{\mathtt{x}}}-1,{{\mathtt{y}}}})+\mathcal{G}_{{{\mathtt{x}}},{{\mathtt{y}}}+1}+\mathcal{G}_{{{\mathtt{x}}},{{\mathtt{y}}}-1}-4\mathcal{G}_{{{\mathtt{x}}},{{\mathtt{y}}}}+{\omega}^2\mathcal{G}_{{{\mathtt{x}}},{{\mathtt{y}}}}=0,\quad ({{\mathtt{x}}}, {{\mathtt{y}}})\in\mathring{\mathbb{H}},
\end{split}\end{equation}
\begin{align}
\label{eqsurfG}
{\alpha}_{{{s}}} \left(\mathcal{G}_{{{\mathtt{x}}}+1,{0}}+\mathcal{G}_{{{\mathtt{x}}}-1,{0}}-2\mathcal{G}_{{{\mathtt{x}}},{0}}\right)+\mathcal{G}_{{{\mathtt{x}}},{}-1}-\mathcal{G}_{{{\mathtt{x}}},{0}}+{\omega}^2{\mathit{m}}_{{{s}}}\mathcal{G}_{{{\mathtt{x}}},{0}}=\delta_{{{\mathtt{x}}}, 0}, \quad({{\mathtt{x}}}, {0})\in\partial{\mathbb{H}}.
\end{align}
\label{discGreenalleq}
\end{subequations}
The solution of above system of equations \eqref{discGreenalleq} can be found in terms of a contour integral as the function $\mathcal{G}:\mathbb{H}\to\mathbb{C}$ given by
\begin{equation}\begin{split}
\mathcal{G}_{{\mathtt{x}},{\mathtt{y}}}=\frac{1}{2\pi i}\int_{{\mathbb{T}}}\frac{{\lambda}^{-{\mathtt{y}}}({{\mathfrak{z}}}){\mathfrak{z}}^{|{\mathtt{x}}|-1}}{{{\mathfrak{K}}}_{{s}}({{\mathfrak{z}}})}d{\mathfrak{z}}, 
\quad
({{\mathtt{x}}}, {{\mathtt{y}}})\in{\mathbb{H}},
\label{defGfn}
\end{split}\end{equation}
where ${\mathbb{T}}$ denotes the counter-clockwise unit circle contour in the complex plane,
and
\begin{equation}\begin{split}
{{{\mathfrak{K}}}}_{{s}}({{\mathfrak{z}}})&={\lambda}({{\mathfrak{z}}})+{\mathfrak{F}}_{{s}}({{\mathfrak{z}}}),\quad
\lambda({{\mathfrak{z}}})= \frac{{\mathtt{r}}({{\mathfrak{z}}})-{\mathtt{h}}({{\mathfrak{z}}})}{{\mathtt{r}}({{\mathfrak{z}}})+{\mathtt{h}}({{\mathfrak{z}}})},\label{lambda}\\
{\mathtt{h}}({{\mathfrak{z}}})&=\sqrt{{\mathtt{Q}}({{\mathfrak{z}}})-2},\quad {\mathtt{r}}({{\mathfrak{z}}})=\sqrt{{\mathtt{Q}}({{\mathfrak{z}}})+2},\quad {\mathtt{Q}}({{\mathfrak{z}}})=4-{{\mathfrak{z}}}-{{\mathfrak{z}}}^{-1}-({{{\omega}}}+i\varepsilon)^2,\\
\text{and }{\mathfrak{F}}_{{s}}({{\mathfrak{z}}})&=m_{{s}}{{{\omega}}}^2-1+\alpha_{{s}}({{\mathfrak{z}}}+{{\mathfrak{z}}}^{-1}-2),
\end{split}\end{equation}
with $\varepsilon>0$ representing a vanishingly small absorption \cite{sK}.
See Section 7(a) in \cite{Blssurf} for few more details regarding the Green's function;
see also Section 4(c) in \cite{Blssurf} and Section 2.2 in \cite{sK} regarding some definitions that appear in \eqref{lambda}.

\begin{remark}
 Note that $\mathcal{G}_{{\mathtt{x}},{\mathtt{y}}}$ decays exponentially as $|{\mathtt{x}}|, |{\mathtt{y}}|\to\infty$ on $\mathbb{H}$ due to the assumption $\varepsilon>0.$
As $\varepsilon\to0+$, the scattering solution is recovered that satisfies radiation conditions \cite{Shaban} in the present two dimensional case; see also \cite{NBls}, where the choice of sign in the imaginary part of the spectral parameter in the resolvent of the discrete Laplacian operator is clarified.
\label{remeps}
\end{remark}

Let
\begin{equation}\begin{split}
\label{zincw}
{\mathfrak{z}}_{\sw}:=e^{i{\mathit k}_{\sw}}, 
\end{split}\end{equation}
where ${\mathit k}_{\sw}$ is incident surface wave number in \eqref{incwave0}. Due to the presence of absorption $\varepsilon>0$, ${\mathfrak{z}}_{\sw}$ lies inside the unit disk in complex plane $\mathbb{C}$, i.e. $|{\mathfrak{z}}_{\sw}|<1$. Moreover, according to the definition in \eqref{lambda}, $\lambda({{\mathfrak{z}}}_{\sw})=\lambda({{\mathfrak{z}}}_{\sw}^{-1})=e^{-{\eta}({{\mathit k}}_{\sw})}$.

\begin{remark}
Note that 
${\mathfrak{F}}_{{s}}({{\mathfrak{z}}})={\mathfrak{F}}_{{s}}({{\mathfrak{z}}}^{-1})$ and $\lambda({{\mathfrak{z}}})=\lambda({{\mathfrak{z}}}^{-1})$ so that
${{{\mathfrak{K}}}}_{{s}}({{\mathfrak{z}}})={{{\mathfrak{K}}}}_{{s}}({{\mathfrak{z}}}^{-1})$.
Also note that ${\widehat{\su}}_{{{\mathtt{x}}}, {0}}={\hat{\rm a}}
{\mathfrak{z}}_{\sw}^{{\mathtt{x}}}$ according to \eqref{incwave0}, \eqref{zincw}, and, in general, ${\widehat{\su}}_{{{\mathtt{x}}}, {{\mathtt{y}}}}={{\hat{\rm a}}}
{\mathfrak{z}}_{\sw}^{{\mathtt{x}}}\lambda^{-{\mathtt{y}}}({{\mathfrak{z}}}_{\sw})$ using the definition in \eqref{lambda}.
\label{remarkKer}
\end{remark}

\begin{remark}
In connection with the eigenfunctions stated earlier, specially Eqs. \eqref{rhogm0}, \eqref{rhogm}, note that
\begin{equation}\begin{split}
\rho(\gamma_{{\sw}})
=\frac{({{\mathfrak{z}}}_{{\sw}}-{{\mathfrak{z}}}_{{\sw}}^{-1})}{{{\mathfrak{z}}}_{{\sw}}{{\mathfrak{K}}}_{{s}}'({{\mathfrak{z}}}_{{\sw}})},\quad
\rho(\gamma) 
=\frac{1}{\pi}\frac{\Im\lambda_\gamma}{|{\mathfrak{K}}_{{s}}({{\mathfrak{z}}}_\gamma)|^2}, 
\label{JGr2}
\end{split}\end{equation}
where $\lambda_\gamma\equiv\lambda({\mathfrak{z}}_\gamma)$ using the following mappings:
\begin{equation}\begin{split}
\gamma\equiv\gamma({\mathfrak{z}})=2-{\mathfrak{z}}-{\mathfrak{z}}^{-1} \text{ and }{\mathfrak{z}}_{\gamma}\equiv{\mathfrak{z}}(\gamma)\text{ such that }|{\mathfrak{z}}_{\gamma}|<1.
\end{split}\end{equation}
In addition to Remark \ref{remonedim}, note that ${\mathfrak{z}}_{\gamma_{{\sw}}}={\mathfrak{z}}_0$ defined in \eqref{zincw} while $\gamma({\mathfrak{z}}_{{\sw}})=\gamma_{{\sw}}$.
In this context, see Appendix \ref{app:secGreen} as well.
\label{remrholam}
\end{remark}

In the backdrop of the Green's function described above in \eqref{defGfn},
hence,
the solution of \eqref{eqbulk} and \eqref{eqsurf1}, that decays exponentially as $|{\mathtt{x}}|, |{\mathtt{y}}|\to\infty$ on $\mathbb{H}$ under the assumption $\varepsilon>0,$ 
can be expressed, in terms of the point sources $\{f_j\}_{j=1}^{\Ld}$ \eqref{deffj}, as
\begin{equation}\begin{split}
{{\mathit{u}}}_{{{\mathtt{x}}}, {{\mathtt{y}}}}=\sum_{j=1}^{\Ld}\mathcal{G}_{|{\mathtt{x}}-j|,{\mathtt{y}}}f_j,\quad ({{\mathtt{x}}}, {{\mathtt{y}}})\in{\mathbb{H}}.
\label{soluxy}
\end{split}\end{equation}
However, the definition \eqref{deffj} includes the scattered field $\{{{\mathit{u}}}_{{{\mathtt{x}}}, {0}}\}_{{\mathtt{x}}=1}^{\Ld}$ on the perturbed sites.
By restricting the solution ${{\mathit{u}}}$ \eqref{soluxy} to $\partial\mathbb{H}$, we have 
\begin{equation}\begin{split}
{{\mathit{u}}}_{{{\mathtt{x}}}, {0}}
=\sum_{j=1}^{\Ld}\mathcal{G}_{|{\mathtt{x}}-j|,0}f_j,\quad {\mathtt{x}}\in\mathbb{Z},
\label{eqsurf2}
\end{split}\end{equation}
where according to \eqref{defGfn}
\begin{equation}\begin{split}
\mathcal{G}_{|{\mathtt{x}}|,0}=\frac{1}{2\pi i}\int_{{\mathbb{T}}}\frac{{\mathfrak{z}}^{|{\mathtt{x}}|-1}}{{{\mathfrak{K}}}_{{s}}({{\mathfrak{z}}})}d{\mathfrak{z}}.
\label{Green0}
\end{split}\end{equation}
Above equation \eqref{eqsurf2} for ${\mathtt{x}}\in[1,\Ld]\cap\mathbb{Z}$ yields a closed system of linear algebraic equations for
$\{{{\mathit{u}}}_{{{\mathtt{x}}}, {0}}\}_{{\mathtt{x}}=1}^{\Ld}$ so that the problem is reduced to the inversion of a $\Ld\times \Ld$ 
matrix.

\begin{remark}
Consider the following Fourier transforms:
\begin{equation}\begin{split}
f({\mathfrak{z}})=\sum\nolimits_{j=1}^{\Ld}{\mathfrak{z}}^{-j}f_{j},
\quad
{{\mathit{u}}}_0^{\rm F}({{\mathfrak{z}}})=\sum\nolimits_{{\mathtt{x}}\in\mathbb{Z}}{{\mathit{u}}}_{{\mathtt{x}},0}{\mathfrak{z}}^{-{\mathtt{x}}},
\quad
{{\mathcal{G}}}_0^{\rm F}({{\mathfrak{z}}})=\sum\nolimits_{{\mathtt{x}}\in\mathbb{Z}}{\mathcal{G}}_{{\mathtt{x}},0}{\mathfrak{z}}^{-{\mathtt{x}}},
\label{deffjFT}
\end{split}\end{equation}
using \eqref{deffj} in the first expression
and
\eqref{Green0} in the third.
Using \eqref{eqsurf2} and the Fourier transforms
$f, {{\mathit{u}}}_0^{\rm F}$, $\mathcal{G}_0^{\rm F}$ defined in \eqref{deffjFT}, it follows that
\begin{equation}\begin{split}
{{\mathfrak{K}}}_{{s}}({{\mathfrak{z}}}){{\mathit{u}}}_0^{\rm F}({{\mathfrak{z}}})=f({\mathfrak{z}}),
\label{solu0}
\end{split}\end{equation}
as ${{{\mathfrak{K}}}}_{{s}}({{\mathfrak{z}}}){\mathcal{G}}^{\rm F}_0({{\mathfrak{z}}})=1$.
Further, by Remark \ref{remarkKer}, ${\mathcal{G}}^{\rm F}_0$ satisfies the property
\begin{equation}\begin{split}
{\mathcal{G}}^{\rm F}_0({{\mathfrak{z}}})={\mathcal{G}}^{\rm F}_0({{\mathfrak{z}}^{-1}}).
\label{eqGz}
\end{split}\end{equation}
\label{remarkGFT}
\end{remark}
Using the expression \eqref{Green0} for the Green's function evaluated on $\partial\mathbb{H}$,
and the definition \eqref{deffj}, the scattered field, as given by \eqref{eqsurf2},
on the perturbed sites can be concisely, but formally, solved to obtain
\begin{equation}\begin{split}
\boldsymbol{u}=(\boldsymbol{I}-\boldsymbol{T}\boldsymbol{D}({{\upmu}}))^{-1} \boldsymbol{T}\boldsymbol{D}({{\upmu}})\widehat{\boldsymbol{u}},
\label{uvec0}
\end{split}\end{equation}
where
\begin{subequations}
 \begin{equation}\begin{split}
\boldsymbol{T}=\text{Toeplitz}(\mathcal{G}_{0,0},\mathcal{G}_{1,0},\mathcal{G}_{2,0},\dotsc,\mathcal{G}_{{\Ld-1},0})\in\mathbb{C}^{\Ld\times\Ld},
\end{split}\end{equation}
\begin{equation}\begin{split}
\boldsymbol{D}({{\upmu}})=-m_{{s}}{\omega}^2\text{diag}({{{\upmu}}}_1,{{{\upmu}}}_2,\dotsc,{{{\upmu}}}_{\Ld})\in\mathbb{C}^{\Ld\times\Ld},
\end{split}\end{equation}
\begin{equation}\begin{split}
\boldsymbol{u}=({{\mathit{u}}}_{1,0}, {{\mathit{u}}}_{2,0}, \dotsc, {{\mathit{u}}}_{\Ld,0})\in\mathbb{C}^{\Ld},\quad \widehat{\boldsymbol{u}}={{\hat{\rm a}}}~{\mathfrak{z}}_{\sw}(1,{\mathfrak{z}}_{\sw},{\mathfrak{z}}_{\sw}^2,\dotsc,{\mathfrak{z}}_{\sw}^{\Ld-1})\in\mathbb{C}^{\Ld},
\label{uuinc}
\end{split}\end{equation}
\begin{equation}\begin{split}
\textrm{and }
\boldsymbol{z}={\mathfrak{z}}(1,{\mathfrak{z}},{\mathfrak{z}}^2,\dotsc,{\mathfrak{z}}^{\Ld-1})\in\mathbb{C}^{\Ld}.
\end{split}\end{equation}
\end{subequations}
Using above result \eqref{uvec0} componentwise in \eqref{deffj}, 
the complete solution for the scattered field on $\mathbb{H}$ is given by \eqref{soluxy}.

\subsection{Reflection and transmission coefficients for surface wave incidence}

For ${\mathtt{x}}\in\mathbb{Z}^-,$ as $\varepsilon\to0+$ and ${\mathtt{x}}\to-\infty,$
according to \eqref{eqsurf2},
\begin{align}
    {{\mathit{u}}}_{{\mathtt{x}},0}=-m_{{s}}{\omega}^2\sum_{j=1}^{{\Ld}}{{{\upmu}}}_j\mathcal{G}_{-{\mathtt{x}}+j,0}({{\mathit{u}}}_{j,0}+{\widehat{\su}}_{j,0})=R{\hat{\rm a}}{\mathfrak{z}}_{\sw}^{-{\mathtt{x}}}.
    \end{align}
Hence, using the Cauchy residue Theorem to evaluate \eqref{Green0} considering the fact that ${\mathfrak{z}}_{\sw}<1$ for $\varepsilon>0$, the reflection coefficient is given by
\begin{equation}\begin{split}
R&=-\frac{m_{{s}}{\omega}^2}{{\hat{\rm a}}}(\frac{1}{{\mathfrak{z}}{{\mathfrak{K}}}'_{{s}}({{\mathfrak{z}}})})|_{{\mathfrak{z}}={\mathfrak{z}}_{\sw}}\sum_{j=1}^{{\Ld}}{{{\upmu}}}_j{\mathfrak{z}}_{\sw}^{j}({{\mathit{u}}}_{j}+{\widehat{\su}}_{j})\\
&=\frac{1}{{\hat{\rm a}}}(\frac{1}{{\mathfrak{z}}{{\mathfrak{K}}}'_{{s}}({{\mathfrak{z}}})})|_{{\mathfrak{z}}={\mathfrak{z}}_{\sw}}{\mathfrak{z}}_{\sw}(1,{\mathfrak{z}}_{\sw},{\mathfrak{z}}_{\sw}^2,\dotsc,{\mathfrak{z}}_{\sw}^{\Ld-1})\cdot\boldsymbol{D}({{\upmu}})(\boldsymbol{u}+\widehat{\boldsymbol{u}}),
\label{Rexp1}
\end{split}\end{equation}
where the second line uses \eqref{uuinc}${}_2$.

For ${\mathtt{x}}>\Ld,$ as $\varepsilon\to0+$, ${\mathtt{x}}\to+\infty,$
${{\mathit{u}}}_{{\mathtt{x}},0}=-m_{{s}}{\omega}^2\sum_{j=1}^{{\Ld}}{{{\upmu}}}_j\mathcal{G}_{{\mathtt{x}}-j,0}({{\mathit{u}}}_{j}+{\widehat{\su}}_{j}),$
which in turn implies
that the total field behaves as
\begin{align}
    {{\mathit{u}}}_{{\mathtt{x}},0}+\widehat{\su}_{{\mathtt{x}},0}
={\widehat{\su}}_{{\mathtt{x}},0}+(-m_{{s}}{\omega}^2)\sum_{j=1}^{{\Ld}}{{{\upmu}}}_j\mathcal{G}_{{\mathtt{x}}-j,0}({{\mathit{u}}}_{j}+{\widehat{\su}}_{j})
=T{\hat{\rm a}}{\mathfrak{z}}_{\sw}^{{\mathtt{x}}}.
\end{align}
Simplifying the expression in the same way as that of the reflection coefficient $R$ \eqref{Rexp1}, we find that the transmission coefficient is, therefore, given by
\begin{equation}\begin{split}
T&=1+\frac{(-m_{{s}}{\omega}^2)}{{\hat{\rm a}}}(\frac{1}{{\mathfrak{z}}{{\mathfrak{K}}}'_{{s}}({{\mathfrak{z}}})})|_{{\mathfrak{z}}={\mathfrak{z}}_{\sw}}\sum_{j=1}^{{\Ld}}{{{\upmu}}}_j{\mathfrak{z}}_{\sw}^{-j}({{\mathit{u}}}_{j}+{\widehat{\su}}_{j})\\
&=1+\frac{1}{{\hat{\rm a}}}(\frac{1}{{\mathfrak{z}}{{\mathfrak{K}}}'_{{s}}({{\mathfrak{z}}})})|_{{\mathfrak{z}}={\mathfrak{z}}_{\sw}}{\mathfrak{z}}_{\sw}^{-1}(1,{\mathfrak{z}}^{-1}_{\sw},{\mathfrak{z}}_{\sw}^{-2},\dotsc,{\mathfrak{z}}_{\sw}^{-\Ld+1})\cdot\boldsymbol{D}({{\upmu}})(\boldsymbol{u}+\widehat{\boldsymbol{u}}).
\label{Texp1}
\end{split}\end{equation}

With $\widehat{\boldsymbol{u}}={\hat{\rm a}}{\mathfrak{z}}_{\sw}(1,{\mathfrak{z}}_{\sw},{\mathfrak{z}}_{\sw}^2,\dotsc,{\mathfrak{z}}_{\sw}^{\Ld-1})$ as in \eqref{uuinc}${}_2$,
using \eqref{uvec0},
the reflection coefficient $R$ \eqref{Rexp1} can be succintly expressed as
\begin{equation}\begin{split}
R&=\frac{1}{{\hat{\rm a}}^2}(\frac{1}{{\mathfrak{z}}{{\mathfrak{K}}}'_{{s}}({{\mathfrak{z}}})})|_{{\mathfrak{z}}={\mathfrak{z}}_{\sw}}\boldsymbol{D}({{\upmu}})\widehat{\boldsymbol{u}}\cdot(\boldsymbol{I}-\boldsymbol{T}\boldsymbol{D}({{\upmu}}))^{-1}\widehat{\boldsymbol{u}},
\label{Rexpshort}
\end{split}\end{equation}
while
the transmission coefficient $T$ \eqref{Texp1} is given by
\begin{equation}\begin{split}
T
&=1+\frac{1}{|{\hat{\rm a}}|^2}(\frac{1}{{\mathfrak{z}}{{\mathfrak{K}}}'_{{s}}({{\mathfrak{z}}})})|_{{\mathfrak{z}}={\mathfrak{z}}_{\sw}}\boldsymbol{D}({{\upmu}})\overline{\widehat{\boldsymbol{u}}}\cdot(\boldsymbol{I}-\boldsymbol{T}\boldsymbol{D}({{\upmu}}))^{-1}\widehat{\boldsymbol{u}}.
\label{Texpshort}
\end{split}\end{equation}
The expressions \eqref{Rexpshort}, \eqref{Texpshort} are utilized later in the article for numerical evaluations and graphical illustrations.
At this point it is clear that, using the coefficients \eqref{Rexp1}, \eqref{Texp1}, one can determine the energy flux in the surface waves reflected by the patch of size $\Ld$ of perturbed masses and transmitted across it. More importantly, what remains to be proven is the fact that the all propagative waves, including waves excited in the bulk lattice, carry energy flux that equals that of incident wave. This task is carried out next.

\subsection{Energy conservation} 

The energy flux in the surface wave \eqref{surfwavemode} is \cite{Brillouin}
\begin{equation}
\begin{split}
\mathcal{E}({{{\mathit k}}_{\sw}})=\frac{1}{2}{\omega}^2|{u}_{{0},{0}}|^2\mu_{{{s}}}({{\mathit k}}_{\sw}){\mathit v}_{{{s}}}({{{\mathit k}}_{\sw}}),\\ \textrm{with }\mu_{{{s}}}({{\mathit k}}_{\sw}):=({\mathit{m}}_{{{s}}}-1)+(1-e^{-2{{{\eta}({{{\mathit k}}_{\sw}})}}})^{-1},
\label{muSwave}
\end{split}
\end{equation}
where ${\mathit v}_{{{s}}}$ is defined in \eqref{surfgpvel}.
The effective mass $\mu_{{{s}}}$ satisfies $\mu_{{{s}}}({{{\mathit k}}_{\sw}})=\mu_{{{s}}}(-{{{\mathit k}}_{\sw}})$ as ${\eta}({{{\mathit k}}_{\sw}})={\eta}(-{{{\mathit k}}_{\sw}})$ while ${\mathit v}_{{{s}}}({{{\mathit k}}_{\sw}})=-{\mathit v}_{{{s}}}(-{{{\mathit k}}_{\sw}})$.
Using the expression \eqref{muSwave}, the incident surface wave carries the energy flux \cite{Blssurf}
\begin{align}
\widehat{\mathcal{E}}={\mathcal{E}}({\mathit k}_{\sw}),
\label{muSwave0}
\end{align}
with $\widehat{\su}_{{0},{0}}={{\hat{\rm a}}}$ replacing ${u}_{{0},{0}}$.
The expression of energy flux is obtained, as in Appendix 1 of \cite{Blssurf}, from the standard
the definition \cite{Brillouin}
\begin{equation}\begin{split}
\widehat{\mathcal{E}}&:=\lim_{{\mathtt{x}}\to-\infty}\frac{1}{2}\Re\sum_{{\mathtt{y}}\in\mathbb{Z}^-\cup\{0\}}\widehat{\mathtt{f}}_{{\mathtt{x}},{\mathtt{y}}}\overline{\dot{\widehat{\su}}_{{\mathtt{x}},{\mathtt{y}}}},
\label{Einc}
\end{split}\end{equation}
where
$\widehat{\mathtt{f}}_{{\mathtt{x}},{\mathtt{y}}}$ denotes the force on particle at $({{\mathtt{x}},{\mathtt{y}}})$ from the interaction with $({{\mathtt{x}}-1,{\mathtt{y}}})$
while $\dot{\widehat{\su}}_{{\mathtt{x}},{\mathtt{y}}}$ denotes the velocity of the particle at $({{\mathtt{x}},{\mathtt{y}}})$; indeed, $\dot{\widehat{\su}}_{{\mathtt{x}},{\mathtt{y}}}=-i{\omega}{\widehat{\su}}_{{\mathtt{x}},{\mathtt{y}}}$.
The expression \eqref{muSwave}, \eqref{muSwave0} can be simplified for the incident surface wave, as in \eqref{incwave0} with amplitude ${{\hat{\rm a}}}$, 
to the form
\begin{align}
\widehat{\mathcal{E}}=&\frac{1}{2}|{{\hat{\rm a}}}|^2{\omega}\frac{\sin{\mathit k}_{\sw}}{\rho({\gamma_{\sw}})},
\label{Einc1}
\end{align}
where, using \eqref{zincw} and Remark \ref{remrholam},
\begin{equation}\begin{split}
\frac{1}{\rho({\gamma_{\sw}})}
=\frac{{{\mathfrak{z}}}_{\sw}{{\mathfrak{K}}}_{{s}}'({{\mathfrak{z}}}_{\sw})}{({{\mathfrak{z}}}_{\sw}-{{\mathfrak{z}}}_{\sw}^{-1})}
=\frac{|{{\mathfrak{K}}}_{{s}}'({{\mathfrak{z}}}_{\sw})|}{2\sin{\mathit k}_{\sw}}.
\label{rgam0}
\end{split}\end{equation}
Under the assumption of point mass perturbation of finite size, we observe that the transmitted surface wave ${{\mathit{u}}}^{T}$ has the same form as the incident wave $\widehat{\su}$ and same wave number ${\mathit k}_{\sw}$. Therefore, analogous to \eqref{Einc}, the energy flux carried by the transmitted surface wave, with amplitude $T$, is
\begin{align}
\mathcal{E}_{T}=&\frac{1}{2}|{{\hat{\rm a}}}|^2|T|^2{\omega} \frac{\sin{\mathit k}_{\sw}}{\rho({\gamma_{\sw}})}.
\label{Etran1}
\end{align}
We also observe that the reflected surface wave ${{\mathit{u}}}^{R}$ has the same form as the incident wave $\widehat{\su}$ but wave number $-{\mathit k}_{\sw}$; in the complex plane the corresponding point is ${\mathfrak{z}}_{\sw}^{-1}$ and ${{\mathit{u}}}^{R}_{{\mathtt{x}},0}={{\hat{\rm a}}}R{\mathfrak{z}}_{\sw}^{-{\mathtt{x}}}$ (recall the discussion in the context of \eqref{Rexp1}). Thus, the energy flux in the reflected surface wave, with amplitude $R$,
is given by
\begin{equation}\begin{split}
\mathcal{E}_{R}&=\frac{1}{2}|{{\hat{\rm a}}}|^2|R|^2{\omega} \frac{\sin{\mathit k}_{\sw}}{\rho({\gamma_{\sw}})}.
\label{Eref1}
\end{split}\end{equation}

\begin{figure}[htb!]
\centering
\includegraphics[width=.4\textwidth]{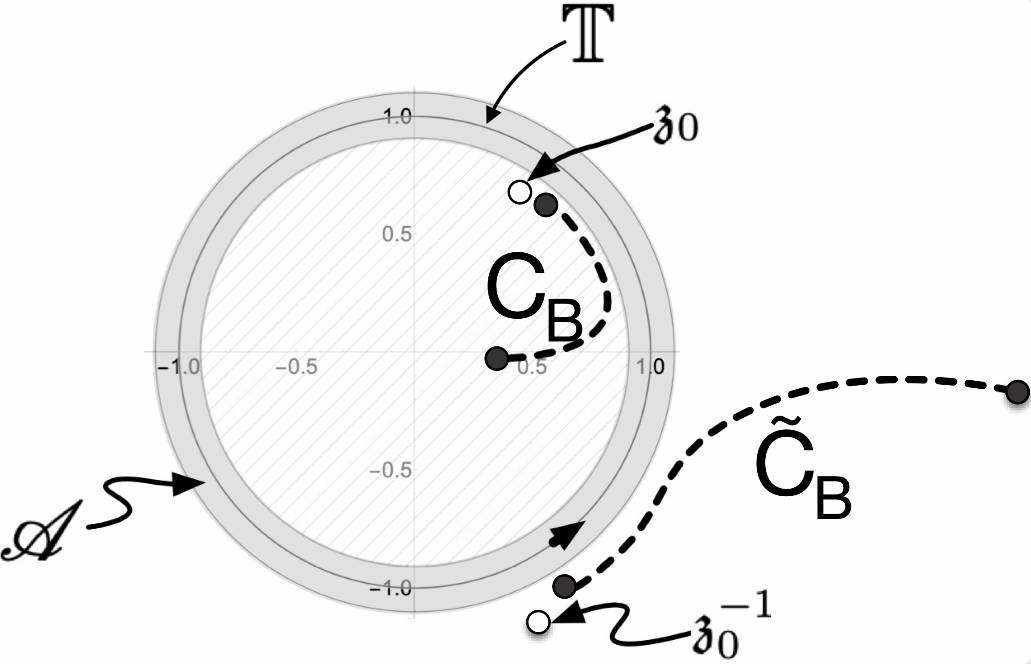}
\caption{Schematic of $C_B$ in the complex plane $\mathbb{C}$.
$\tilde{C}_B$ is related to $C_B$ by the mapping ${\mathfrak{z}}\mapsto1/{\mathfrak{z}}.$
The thickness of white annulus region containing the unit circle depends on $\varepsilon$ and shrinks as $\varepsilon\to0+$.}
\label{branchcuthalfplane}
\end{figure}

The energy leaked into the half space is
\begin{equation}\begin{split}
\mathcal{E}_{\mathring{\mathbb{H}}}
&=-\lim_{{\mathtt{y}}\to-\infty}\frac{1}{2}\Re\sum_{{\mathtt{x}}\in\mathbb{Z}}\mathtt{f}_{{\mathtt{x}},{\mathtt{y}}}\dot{{{\mathit{u}}}}_{{\mathtt{x}},{\mathtt{y}}},
\label{Ehalf}
\end{split}\end{equation}
where $\mathtt{f}_{{\mathtt{x}},{\mathtt{y}}}$ denotes the force on particle at $({{\mathtt{x}},{\mathtt{y}}})$ from the interaction with $({{\mathtt{x}},{\mathtt{y}}}-1)$ while taking into account the scattered wave field ${{\mathit{u}}}$ on $\mathbb{H}$.
With $\varepsilon=\Im{\omega}\to0+,$
using the Plancherel theorem with the Fourier transforms \eqref{deffjFT}, and the property \eqref{eqGz}, we find that the definition \eqref{Ehalf} leads to the expression
\begin{equation}\begin{split}
\mathcal{E}_{\mathring{\mathbb{H}}}
=&\frac{1}{2}{{\omega}}\frac{1}{2\pi i}\oint_{{C}_B}\Im\lambda({\mathfrak{z}})\frac{1}{2}(|{{\mathit{u}}}_0^{\rm F}({\mathfrak{z}})|^2+|{{\mathit{u}}}_0^{\rm F}({\mathfrak{z}}^{-1})|^2){\mathfrak{z}}^{-1}d{\mathfrak{z}},
\label{Ehalf1}
\end{split}\end{equation}
$\text{ since }|\lambda|=1\text{ on }C_B\text{ and }|\lambda|<1\text{ outside }C_B,$
where
$C_B$ is a part of the branch cut for $\lambda$ inside the unit circle and coincides with the continuous spectrum (propagating and evanescent bulk waves) in ${\mathfrak{z}}$ plane.
See Fig. \ref{branchcuthalfplane} for an exaggerated schematic of $C_B$ assuming $\varepsilon>0$.

\begin{proposition}
We have the conservation of energy relation
\begin{equation}\begin{split}
\mathcal{E}_{R}+\mathcal{E}_{T}+\mathcal{E}_{\mathring{\mathbb{H}}}=\widehat{\mathcal{E}}.
\label{Ebal}
\end{split}\end{equation}
where $\widehat{\mathcal{E}}$ is given by \eqref{Einc}, $\mathcal{E}_{R}$ is given by \eqref{Eref1}, $\mathcal{E}_{T}$ is given by \eqref{Etran1}, and $\mathcal{E}_{\mathring{\mathbb{H}}}$ is given by \eqref{Ehalf1}.
\label{energybal}
\end{proposition}

\begin{proof}
Recall Remark \ref{remarkGFT}.
Using \eqref{solu0}, \eqref{deffjFT}, \eqref{Einc1}, \eqref{Etran1}, \eqref{Eref1}, and \eqref{Ehalf1},
and \eqref{lambda}, the statement of energy balance \eqref{Ebal} becomes
\begin{equation}\begin{split}
\frac{1}{2}{{\omega}}\frac{1}{2\pi i}\oint_{{C}_B}\frac{\Im\lambda_\gamma \frac{1}{2}(|{f}({\mathfrak{z}})|^2+|{f}({\mathfrak{z}}^{-1})|^2)}{({\mathfrak{F}}_{{s}}+\lambda)({\mathfrak{F}}_{{s}}+\lambda^{-1})}{\mathfrak{z}}^{-1}d{\mathfrak{z}}\\
+\frac{1}{2}{{\omega}}\frac{|f({\mathfrak{z}}_{\sw}^{-1})|^2+|f({\mathfrak{z}}_{\sw})|^2}{-2i{{\mathfrak{z}}}_{\sw}{{\mathfrak{K}}}_{{s}}'({{\mathfrak{z}}}_{\sw})}
+\frac{1}{2}{{\omega}}\Im(f({\mathfrak{z}}_{\sw}))=0.
\label{thmeq1}
\end{split}\end{equation}
The first two terms in \eqref{thmeq1} can be combined by an application of Cauchy's residue theorem to re-write \eqref{Ebal} as
\begin{equation}\begin{split}
\frac{1}{2}{{\omega}}\frac{1}{2\pi i}\oint_{{{\mathbb{T}}}}\frac{\Im\lambda_\gamma \frac{1}{2}(|{f}({\mathfrak{z}})|^2+|{f}({\mathfrak{z}}^{-1})|^2)}{({\mathfrak{F}}_{{s}}+\lambda)({\mathfrak{F}}_{{s}}+\lambda^{-1})}{\mathfrak{z}}^{-1}d{\mathfrak{z}}
+\frac{1}{2}{{\omega}}\Im(f({\mathfrak{z}}_{\sw}))=0.
\label{thmeq2}
\end{split}\end{equation}
Consider the expression $W={-\frac{1}{2}}\Re\sum\nolimits_{{\mathtt{x}}=1}^{\Ld} f_{{\mathtt{x}}}\overline{\dot{{{\mathit{u}}}}_{{\mathtt{x}},0}}$ using \eqref{deffj}. By an application of Plancherel theorem
\begin{equation}\begin{split}
W=-\frac{1}{2} {\omega}\frac{1}{2\pi i}\oint_{{\mathbb{T}}}|f({\mathfrak{z}})|^2\frac{1}{2}\Im(\frac{1}{{\mathfrak{K}}({\mathfrak{z}})}+\frac{1}{{\mathfrak{K}}({\mathfrak{z}}^{-1})}){\mathfrak{z}}^{-1}d{\mathfrak{z}},
\end{split}\end{equation}
which upon using the definitions \eqref{lambda}
leads to
\begin{equation}\begin{split}
W=\frac{1}{2}{{\omega}}\frac{1}{2\pi i}\oint_{{{\mathbb{T}}}}\frac{\Im\lambda_\gamma \frac{1}{2}(|{f}({\mathfrak{z}})|^2+|{f}({\mathfrak{z}}^{-1})|^2)}{({\mathfrak{F}}_{{s}}+\lambda)({\mathfrak{F}}_{{s}}+\lambda^{-1})}{\mathfrak{z}}^{-1}d{\mathfrak{z}}.
\label{thmeq3}
\end{split}\end{equation}
On the other hand, using \eqref{deffj} and by the direct substitution of expression of the incident wave, it is found that
\begin{equation}\begin{split}
W=&-\frac{1}{2}{\omega}\Im(f({\mathfrak{z}}_{\sw})).
\label{thmeq4}
\end{split}\end{equation}
The equations \eqref{thmeq3}, \eqref{thmeq4} lead to \eqref{thmeq2}.
\end{proof}

\begin{remark}
The statement of Proposition \ref{energybal} can be equivalently expressed as
\begin{align}
    |R|^2+|T|^2+\Dsf=1,
    \label{eq:energyconservationrelation}
\end{align}
    where ${\Dsf}=\mathcal{E}_{\mathring{\mathbb{H}}}/\widehat{\mathcal{E}}>0$
is the energy flux radiated into the lattice half plane ${\mathring{\mathbb{H}}}$ per unit incident wave energy flux while $|R|^2$ and $|T|^2$ represent the surface wave reflectance and transmittance, respectively.
\end{remark}

\section{Multiscale analysis for small, random mass perturbations}
\label{secmain}
We assume that there is a unit-amplitude forward-going (normalized) surface mode incoming from the left region ${\mathtt{x}}\leq 0$ and radiation condition into the right region ${\mathtt{x}} \geq \Ld+1$.

The total wave field ${\su}$ is solution of \eqref{eqbulk} for $\mathtt{y}\neq 0$,  
\begin{equation}\begin{split}
\alpha_{{s}}({\su}_{{{\mathtt{x}}}+1, {{\mathtt{y}}}}+{\su}_{{{\mathtt{x}}}-1, {{\mathtt{y}}}}-2{\su}_{{{\mathtt{x}}}, {{\mathtt{y}}}})+{\su}_{{{\mathtt{x}}}, {{\mathtt{y}}}-1}-{\su}_{{{\mathtt{x}}}, {{\mathtt{y}}}}+m_{{s}} {\omega}^2{\su}_{{{\mathtt{x}}}, {{\mathtt{y}}}}=0, \quad {\mathtt{y}}=0,
\end{split}\end{equation}
for ${\mathtt{x}} \not\in [1,\Ld]\cap\mathbb{Z}$, and \begin{equation}\begin{split}
\alpha_{{s}}({\su}_{{{\mathtt{x}}}+1, {{\mathtt{y}}}}+{\su}_{{{\mathtt{x}}}-1, {{\mathtt{y}}}}-2{\su}_{{{\mathtt{x}}}, {{\mathtt{y}}}})+{\su}_{{{\mathtt{x}}}, {{\mathtt{y}}}-1}-{\su}_{{{\mathtt{x}}}, {{\mathtt{y}}}}+m_{{s}}(1+{\upmu}_{{\mathtt{x}}}) {\omega}^2{\su}_{{{\mathtt{x}}}, {{\mathtt{y}}}}=0, \quad {\mathtt{y}}=0,
\end{split}\end{equation}
for ${\mathtt{x}} \in [1,\Ld]\cap\mathbb{Z}$.

We denote by $\widehat{\su}$ the propagative surface wave mode that is incident from the left side:
\begin{equation} 
\widehat{\su}_{{{\mathtt{x}}},{{\mathtt{y}}}}=e^{{ i{{\mathit k}}_{\sw} {{\mathtt{x}}}} }\phi_{\gamma_0} ({{\mathtt{y}}}).
\label{incwJG}
\end{equation}
Note that the expression \eqref{incwJG} coincides with that stated in \eqref{incwave0} provided the amplitude is chosen such that ${\hat{\rm a}}=\sqrt{\rho( \gamma_0)}$. 

We denote by ${{\mathit{u}}}= {\su} -\widehat{\su} $, as in Section \ref{secexact}, the scattered field that satisfies the radiating conditions: \\
i) for $\gamma \in (0,{\omega}^2) \cup \{\gamma_0\} $, $\langle {{\mathit{u}}}_{{\mathtt{x}},\cdot},\phi_{\gamma} \rangle\exp(i k(\gamma) {{\mathtt{x}}} ) $ does not depend on ${\mathtt{x}}$ for ${{\mathtt{x}}} \leq 0$ and $\langle{{\mathit{u}}}_{{\mathtt{x}},\cdot},\phi_{\gamma} \rangle\exp(-i k(\gamma) {{\mathtt{x}}} ) $ does not depend on ${\mathtt{x}}$ for ${{\mathtt{x}}} \geq \Ld+1$,\\
ii) for $\gamma \in ({\omega}^2-4, 0)$, $\langle{{\mathit{u}}}_{{\mathtt{x}},\cdot},\phi_{\gamma} \rangle\exp(- k(\gamma) {{\mathtt{x}}} ) $ does not depend on ${\mathtt{x}}$ for ${{\mathtt{x}}} \leq 0$ and $\langle{{\mathit{u}}}_{{\mathtt{x}},\cdot},\phi_{\gamma} \rangle\exp( k(\gamma) {{\mathtt{x}}} ) $ does not depend on ${\mathtt{x}}$ for ${{\mathtt{x}}} \geq \Ld+1$.

We define by $R$ the amplitude of the left-going surface mode in the left region ${\mathtt{x}}\leq 0$ and by $T$ the amplitude of the right-going surface mode in the right region ${\mathtt{x}}\geq \Ld+1$:
\begin{align}
R &= \langle {{\mathit{u}}}_{{\mathtt{x}},\cdot} , \phi_{\gamma_0} \rangle \exp(i k(\gamma_0) {{\mathtt{x}}})
\, \mbox{ for } {{\mathtt{x}}} \leq 0,\\
T&= \langle {{\mathit{u}}}_{{\mathtt{x}},\cdot} , \phi_{\gamma_0} \rangle \exp(-i k(\gamma_0) {{\mathtt{x}}}) \, \mbox{ for } {{\mathtt{x}}} \geq \Ld+1.\end{align}
We refer to $R$, resp. $T$, as the reflection, resp. transmission, coefficient, as in Section~\ref{secexact}.

The total wave field ${\su}$ can be expanded as \eqref{eq:expandsol}.
The modal amplitudes $\check{\su}_\gamma({\mathtt{x}}) = \langle{\su}_{{{\mathtt{x}}},\cdot},\phi_\gamma \rangle$ satisfy the coupled difference equations
\begin{align}
\nonumber
&\check{\su}_\gamma({\mathtt{x}}+1)+\check{\su}_\gamma({\mathtt{x}}-1)+(\gamma-2) \check{\su}_\gamma({\mathtt{x}}) = - m_{{s}} {{\upmu}}_{{\mathtt{x}}} {\omega}^2 \sqrt{\rho(\gamma)}\\
&\quad \times \Big[ \sqrt{\rho(\gamma_0)} \check{\su}_{\gamma_0}({\mathtt{x}}) 
+ 
\int_{{\omega}^2-4}^{{\omega}^2} \sqrt{\rho(\gammap)}
\check{\su}_{\gammap}({\mathtt{x}}) d\gammap \Big],\quad{\mathtt{x}}\in [1,\Ld]. 
\label{eq:cma2}
\end{align}

\subsection{Moments of the reflectance, transmittance, and radiative loss}
We assume that the $({{\upmu}}_{{\mathtt{x}}})_{{\mathtt{x}} \in [1,\Ld]\cap\mathbb{Z}}$ are independent and identically distributed random variables with mean zero and variance $\sigma^2$. 
We here carry out a multiscale analysis in the asymptotic framework $\sigma\ll 1 $ and $\Ld\gg 1 $ such that $\Ld\sigma^2 =O(1)$. We prove, in Section \ref{secmainproof}, the following proposition that is the main result of the paper.
\begin{proposition}
\label{prop:mainasy}
Let us denote
\begin{equation}
\frac{1}{\Ld_{\rm loc}}=\frac{\sigma^2 m_{{s}}^2 {\omega}^4 \rho(\gamma_0)^2}{4\sin^2(k(\gamma_0))},\qquad
\Lambda = \frac{\sigma^2 m_{{s}}^2 {\omega}^4 \rho(\gamma_0)}{2\sin k(\gamma_0)}
\int_{0}^{{\omega}^2} \frac{\rho(\gamma)}{\sin k({\gamma}) }
 d\gamma .
\label{1byLloc}
\end{equation}
We denote by $\mathcal {R}_p^\sigma(\tilde{L})=\mathbb{E}[|R(\Ld=[\tilde{L} \Ld_{\rm loc}])|^{2p}]$, $p\geq 0$, the moments of the reflection 
coefficient for $\Ld=[\tilde{L} \Ld_{\rm loc}]$ (which is of the order of $\sigma^{-2}$).\\
In the limit $\sigma\to 0$, the moments of the reflection 
coefficient $(\mathcal {R}_p^\sigma)_{p \geq 0}$, converge 
to the solution $(\mathcal {R}_p)_{p\geq 0}$ of the system
\begin{equation}
\partial_{\tilde{L}} \mathcal {R}_p = p^2 \big( \mathcal {R}_{p+1}+\mathcal {R}_{p-1} -2\mathcal {R}_p)-2 \tilde{\Lambda} p \mathcal {R}_p,
\end{equation}
starting from $\mathcal {R}_p(\tilde{L}=0) = {\bf 1}_0(p)$, where 
\begin{equation}
 \tilde{\Lambda} = \Lambda \Ld_{\rm loc}= 2 \frac{\sin k(\gamma_0)}{\rho(\gamma_0)} \int_0^{{\omega}^2} 
\frac{\rho(\gamma)}{\sin k(\gamma)}
 d\gamma .
\label{LamLloc}
\end{equation}
\end{proposition}
Similarly, we find that $(\mathbb{E}[|R|^{2p} |T|^2])_{p\geq 0}$ converge to the solution $(\mathcal {T}_p)_{p\geq 0}$ of the system
\begin{equation}
\partial_{\tilde{L}} \mathcal {T}_p = 
\big( (p+1)^2 (\mathcal {T}_{p+1}-\mathcal {T}_p)+ p^2(\mathcal {T}_{p-1} -\mathcal {T}_p)\big)- \tilde\Lambda (2p+1) \mathcal {T}_p,
\end{equation}
starting from $\mathcal {T}_p(\tilde{L}=0) = {\bf 1}_0(p)$,
and $(\mathbb{E}[|R|^{2p} |T|^4])_{p\geq 0}$ converge to the solution $(\mathcal {U}_p)_{p\geq 0}$ of the system
\begin{equation}
\partial_{\tilde{L}} \mathcal {U}_p = \big( (p+2)^2 (\mathcal {U}_{p+1}-\mathcal {U}_p)+ p^2(\mathcal {U}_{p-1} -\mathcal {U}_p)\big) + \big( 2- \tilde\Lambda (2p+2) \big) \mathcal {U}_p,
\end{equation}
starting from $\mathcal {U}_p(\tilde{L}=0) = {\bf 1}_0(p)$,

Following \cite[Section 9.2.2]{book}, the solutions of these systems have probabilistic representations (see Appendix \ref{app:jump} for a brief introduction to jump Markov processes). In particular,
\begin{align}
\label{eq:probaR2}
\mathcal{R}_p (\tilde{L})&= 
\mathbb{E}\Big[ {\bf 1}_{N^R_{{\tldL}}=0} \exp\Big(-2 \tilde{\Lambda} \int_0^{{\tldL}} N^R_x dx \Big) |N^R_0=p\Big] ,\quad p \geq 0,
\end{align}
where $(N^R_x)_{x\geq 0}$ is a jump Markov process with state space $\mathbb{N}$ and with generator 
$\mathcal {L}^R f(n) = n^2 \big( f(n+1)+f(n-1)-2f(n) \big)$.
This expression for $p=1$ gives the expectation of $|R|^2$ and the expressions for $p=1$ and $p=2$ give the variance of $|R|^2$ in the limit $\sigma \to 0$.
Similarly,
\begin{align}
\label{eq:probaT2} 
\mathcal{T}_p (\tilde{L})&= 
\mathbb{E}\Big[ {\bf 1}_{N^T_{{\tldL}}=0} \exp\Big(- \tilde{\Lambda}\int_0^{{\tldL}} 2N^T_x +1dx \Big) |N^T_0=p\Big],\quad p \geq 0,
\end{align}
where $(N^T_x)_{x\geq 0}$ is a jump Markov process with state space $\mathbb{N}$ and with generator 
$\mathcal {L}^T f(n) = (n+1)^2 (f(n+1)-f(n)) + n^2( f(n-1)-f(n)) $.
This expression for $p=0$ gives the expectation of $|T|^2$  in the limit $\sigma \to 0$.
We also have
\begin{align}
\mathcal{U}_p (\tilde{L})
&= \exp\big( 2 {{\tldL}} \big)
\mathbb{E}\Big[ {\bf 1}_{N^U_{{\tldL}}=0} \exp\Big( - \tilde\Lambda \int_0^{{\tldL}} 2N^U_x +2dx \Big) |N^U_0=p\Big],\quad p \geq 0,
\label{eq:probaT4} 
\end{align}
where $(N^U_x)_{x\geq 0}$ is a jump Markov process with state space $\mathbb{N}$ and with generator 
$\mathcal {L}^U f(n) = (n+2)^2 (f(n+1)-f(n)) + n^2( f(n-1)-f(n)) $. 
This expression for $p=0$ gives the second moment of $|T|^2$, which in turn gives the variance of $|T|^2$  in the limit $\sigma \to 0$.
These results can also be exploited to determine the moments of the radiative loss.
Indeed, using the energy conservation relation (\ref{eq:energyconservationrelation}) ${\Dsf} +|R|^2+ |T|^2=1$,
the mean radiative loss is
\begin{equation}
 \mathbb{E}[{\Dsf} ] = 1- \mathbb{E}[|R|^2]- \mathbb{E}[|T|^2],
\end{equation}
and its variance is 
\begin{equation}
{\rm Var}({\mathcal D}) = \mathbb{E}[|R|^4] +2\mathbb{E}[|R|^2 |T|^2]+\mathbb{E}[|T|^4] - \mathbb{E}[|R|^2]^2-2\mathbb{E}[|R|^2]\mathbb{E}[|T|^2]-\mathbb{E}[|T|^2]^2 .
\end{equation}

\subsection{Closed form expressions for two regimes}
It seems that it is not possible to find closed form expressions for $\mathbb{E} [|R|^2] $ and $\mathbb{E} [|T|^2] $ in the limit $\sigma \to 0$.
However, we can study the two regimes $\tilde{L} \ll 1 $ (weakly scattering regime) and $\tilde{L}\gg1 $ (strongly scattering regime) and obtain closed form expressions in these two regimes.\\
{\bf 1)} When $\tilde{L} \ll 1$, we can expand (\ref{eq:probaR2}) and (\ref{eq:probaT2}) to get (in the limit $\sigma \to 0$)
\begin{align}
 \mathbb{E}[|R|^2] &= \tilde{L} -(1+\tilde{\Lambda})\tilde{L}^2 +O(\tilde{L}^3) , \\
 \mathbb{E}[|T|^2] &= 1-(1+\tilde{\Lambda})\tilde{L} +(1+\tilde{\Lambda}+\frac{1}{2} \tilde{\Lambda}^2 )\tilde{L}^2 +O(\tilde{L}^3) , \\\ 
 \mathbb{E}[|R|^4] &= 2\tilde{L}^2 +O(\tilde{L}^3) , \\
 \mathbb{E}[|T|^4] &= 1 - 2(1+\tilde{\Lambda})\tilde{L} +(4+4\tilde{\Lambda}+2\tilde{\Lambda}^2)\tilde{L}^2 +O(\tilde{L}^3) , \\
 \mathbb{E}[|R|^2|T|^2] &= \tilde{L} -(3+2\tilde{\Lambda})\tilde{L} +O(\tilde{L}^3) ,
\end{align}
hence
\begin{align}
 \mathbb{E}[{\Dsf}] &= \tilde{\Lambda} \tilde{L} - \frac{1}{2} \tilde{\Lambda}^2 \tilde{L}^2 +O(\tilde{L}^3) = 
 \tilde{\Lambda} \tilde{L} +O(\tilde{L}^2) ,\\
 {\rm Var}({\Dsf}) &= \tilde{L}^2+O(\tilde{L}^3)= O(\tilde{L}^2).
\end{align}
In particular, we observe that the mean radiative loss is proportional to $\Lambda$ to leading order:
\begin{equation}
 \mathbb{E}[{\Dsf}] \simeq  \tilde{\Lambda} \tilde{L} =\Lambda \Ld,
\end{equation}
and the relative variance of radiative loss is constant:
\begin{equation}
{\rm Var}({\Dsf}) /  \mathbb{E}[{\Dsf}]^2\simeq 1/\tilde{\Lambda}^2.
\end{equation}

\begin{figure}[htb!]
\centering
(a)\includegraphics[width=.45\textwidth]{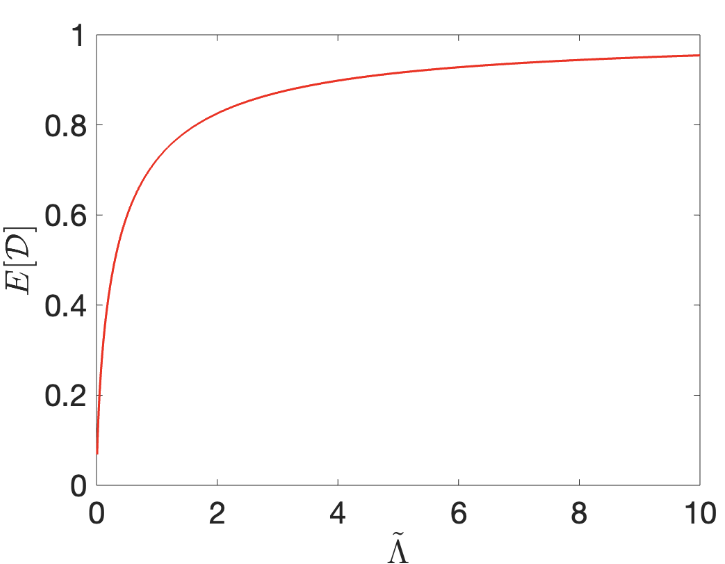} 
(b)\includegraphics[width=.45\textwidth]{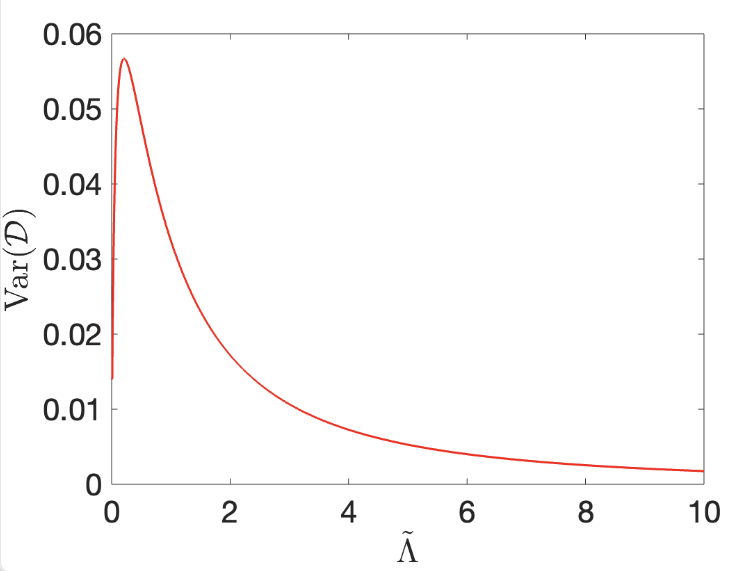}
\caption{
(a) Mean \eqref{ploteq1} and (b) variance \eqref{ploteq2} of the radiative loss $\Dsf$ as functions of $\tilde{\Lambda}$ in the strongly scattering regime $\tilde{L}\gg 1$. 
}
\label{fig_diss_asy}
\end{figure}

\noindent
{\bf 2)} When $\tilde{L} \gg 1$, proceeding as in \cite[Section 9.2.3]{book}, we find $\mathbb{E}[|T|^2] = 0$ and
\begin{equation}
\mathbb{E} [|R|^2] = 1-2 \tilde{\Lambda} \exp(2 \tilde{\Lambda}) E_1 \big( 2\tilde{\Lambda}\big),
\end{equation}
where
\begin{equation}
E_1(x) :=\int_1^\infty \frac{\exp(-xt)}{t} dt
\label{defE1}
\end{equation}
and we remark that 
$\tilde{\Lambda}$
does not depend on $\sigma$. More generally, when $\tilde{L}\gg 1$, we have 
$\mathbb{E} [|T|^{2q} |R|^{2p}]= 0$ if $q\geq 1$ and
\begin{equation}
\mathbb{E} [|R|^{2p}] =\tilde{\Lambda}\int_0^\infty \Big(\frac{s}{2+s}\Big)^p \exp(- \tilde{\Lambda} s) ds,
\end{equation}
for any integer $p$.
In particular, we observe that the mean radiative loss is a monotonically increasing function of $\tilde{\Lambda}$: 
\begin{equation}
 \mathbb{E}[{\Dsf}] = 2 \tilde{\Lambda} \exp(2 \tilde{\Lambda}) E_1 \big( 2\tilde{\Lambda}\big),
 \label{ploteq1}
\end{equation}
where $E_1$ is defined in \eqref{defE1}, and the variance of the radiative loss is a function of $\tilde{\Lambda}$ as well (see Fig.~\ref{fig_diss_asy}):
\begin{equation}
 {\rm Var}({\Dsf}) = \big[ 1-2 \tilde{\Lambda} \exp(2 \tilde{\Lambda}) E_1 \big( 2\tilde{\Lambda}\big) \big]\big[ 1 +2\tilde{\Lambda} + 2 \tilde{\Lambda} \exp(2 \tilde{\Lambda}) E_1 \big( 2\tilde{\Lambda}\big)\big] -1 .
 \label{ploteq2}
\end{equation}

\subsection{Comparison of ensemble of exact solutions vs asymptotics}

\begin{figure}[htb!]
\centering
(a)\includegraphics[width=.45\textwidth]{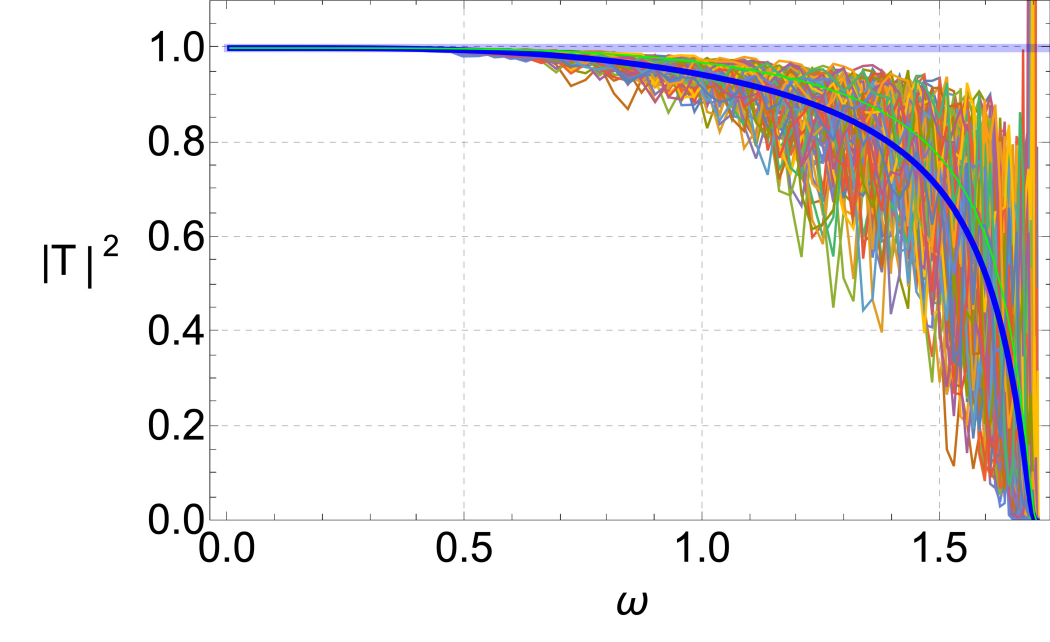}\includegraphics[width=.45\textwidth]{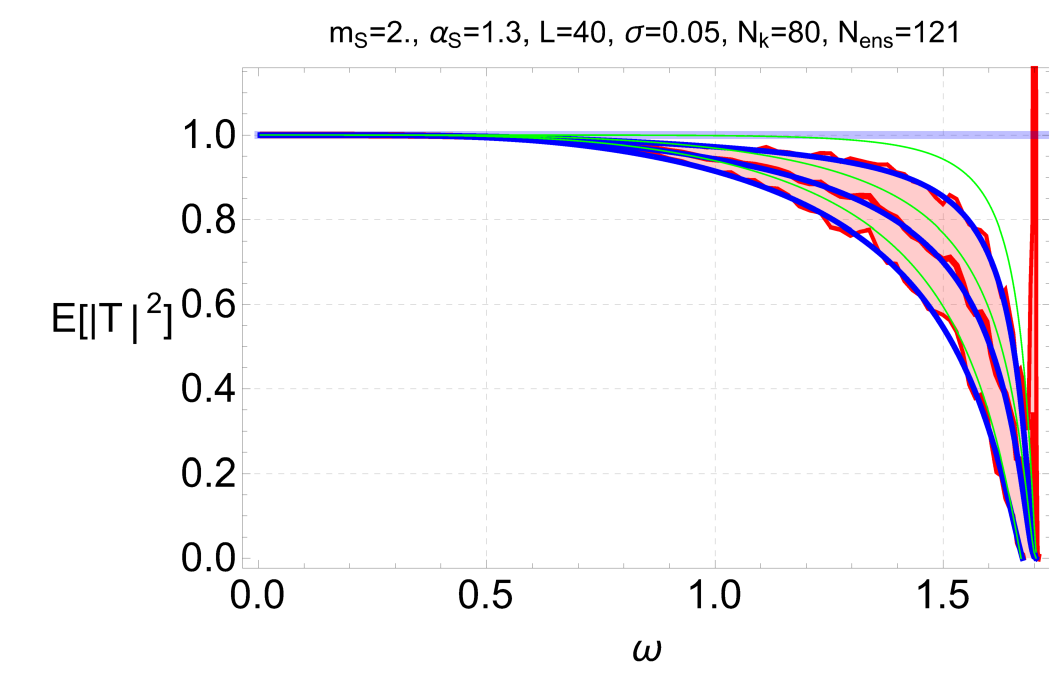}
(b)\includegraphics[width=.45\textwidth]{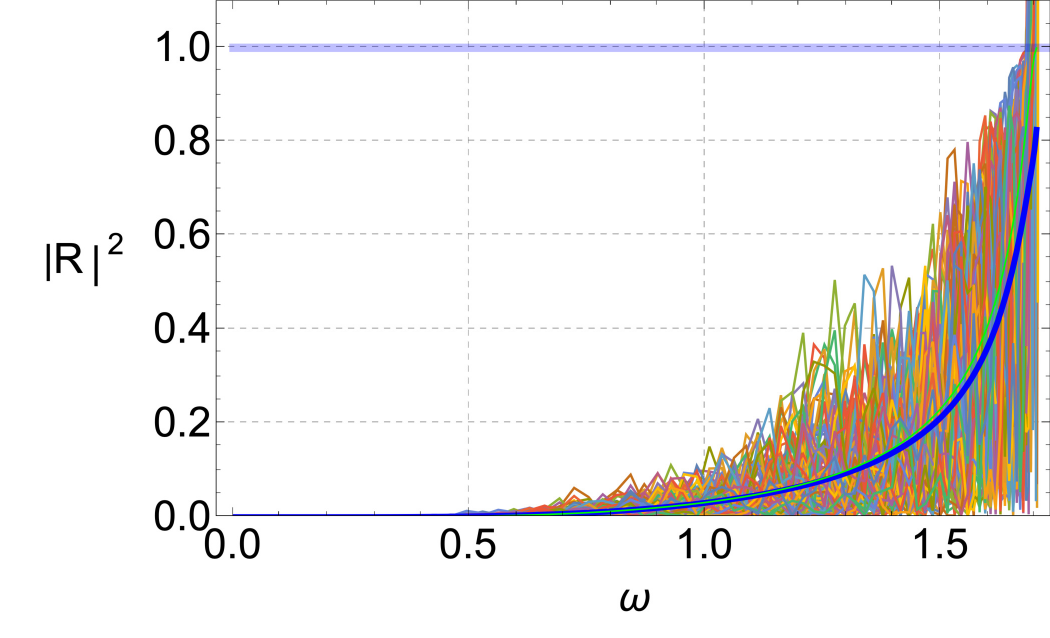}\includegraphics[width=.45\textwidth]{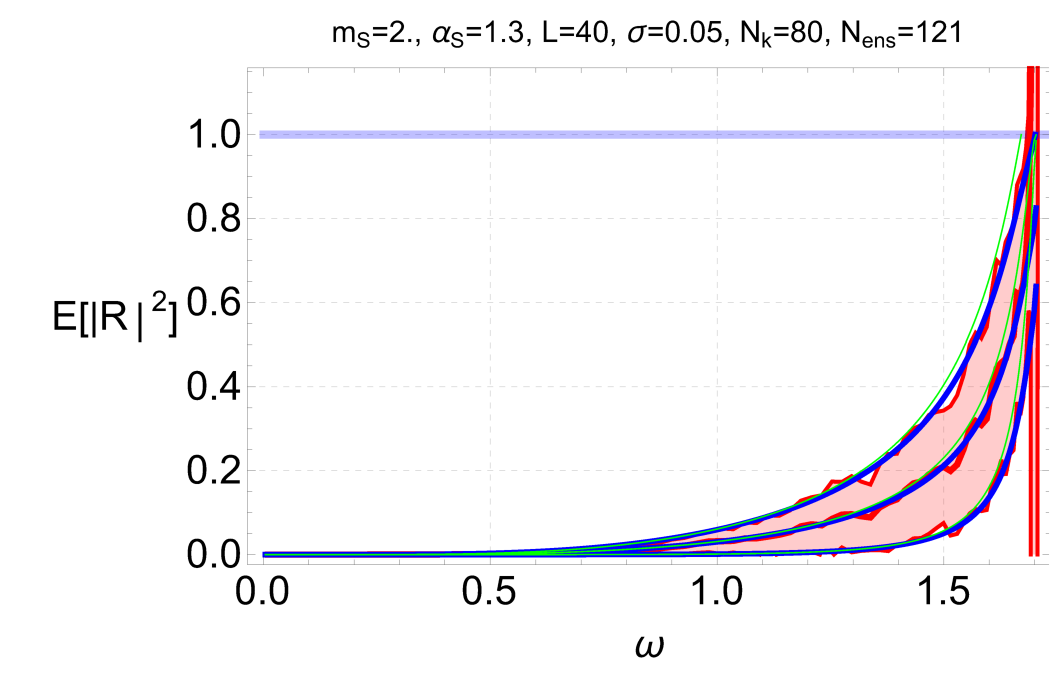}
(c)\includegraphics[width=.48\textwidth]{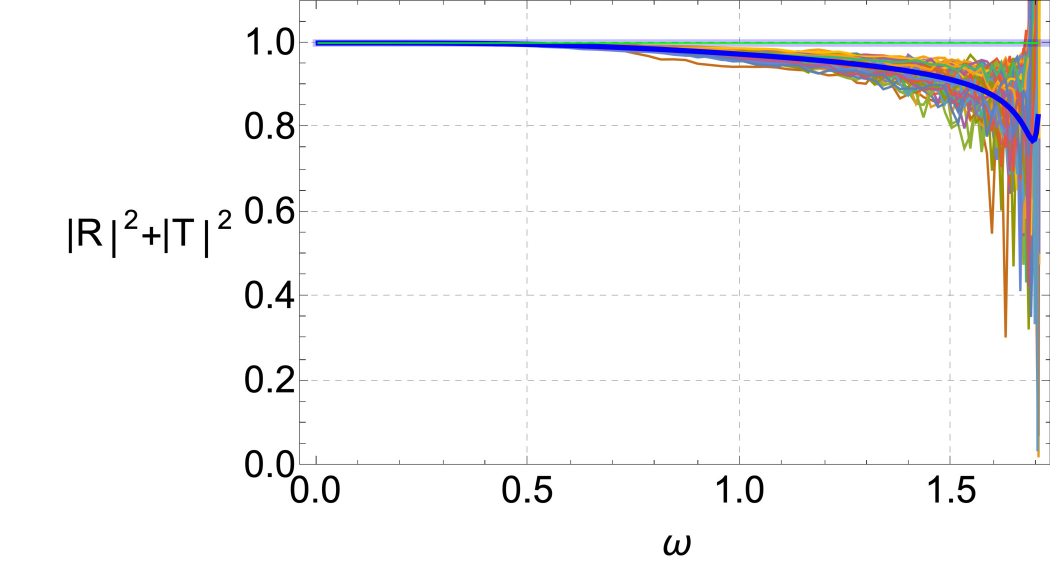}\includegraphics[width=.47\textwidth]{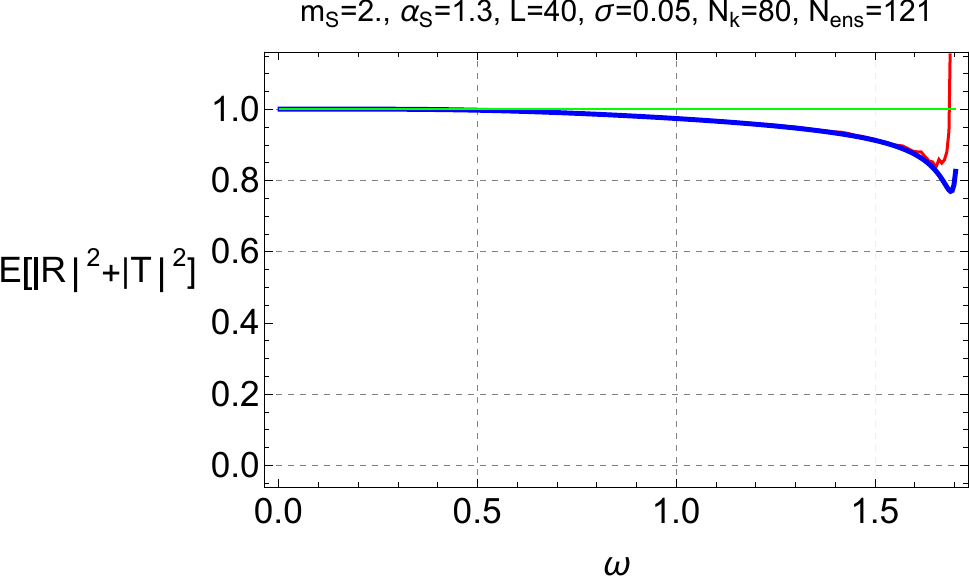}
\caption{
Left: Realizations of (a) $|T|^2$, (b) $|R|^2$, (c) $|R|^2+|T|^2$ using Green's function based exact solution \eqref{Texpshort} and independent and identically distributed mass perturbations.
Right: Expectations and standard deviations of (a) $|T|^2$, (b) $|R|^2$, (c) $|R|^2+|T|^2$.
Blue curves are asymptotic formulas \eqref{eq:probaR2}, \eqref{eq:probaT2}. Red curves are empirical averages using \eqref{Texpshort}.
Green curves represent the theoretical results \eqref{eq:meanR2lambda0} when $\Lambda=0$ (no coupling with radiative modes).
The surface structure parameters $m_s, \alpha_s$ are assumed to be constant outside the perturbed patch $[1,\Ld]$. Here $m_s=2, \alpha_s=1.3$, $\Ld=40$, $\sigma=0.05$.
}
\label{T21asymp}
\end{figure}

In the right of Figure \ref{T21asymp}(a), we compare the empirical averages of numerical simulations (for many different realizations of the random mass perturbations) in the left part of Figure \ref{T21asymp}(a) with the theoretical predictions for the expectation $\mathbb{E}[|T|^2]$ and the standard deviation ${\rm Std}( |T|^{2} )={\rm Var}( |T|^{2} )^{1/2}$. 
The numerical simulations are based on the exact solution \eqref{Texpshort} and the theoretical predictions are based on \eqref{eq:probaT2} and \eqref{eq:probaT4}. 
In a similar manner, in Figure \ref{T21asymp}(b), Figure \ref{T21asymp}(c), we compare the empirical averages of numerical simulations with the theoretical predictions for the expectation $\mathbb{E}[|R|^2]$, $\mathbb{E}[|R|^2+|T|^2]$ and the standard deviation ${\rm Std}( |R|^{2})$. We again obtain an excellent agreement.
We can observe that the behavior of the transmittance close to the right endpoint ${\omega}_{\max}$ (given by \eqref{defwmax}) of the propagative band exhibits large deviations from the perfect transmission. The transmittance goes to one as ${\omega} \to 0$ and to zero as $\omega \to {\omega}_{\max}$. The numerical calculations of the exact solution for $R$ and $T$ for ${\omega}$ very close to ${\omega}_{\max}$ suffer from small errors associated with the evaluation of contour integral and vanishing of group velocity of the surface wave; this can be rectified by employing refined techniques but we did not pursue that as it is a very narrow regime and our focus remains on asymptotic results.
Finally, the green curves in figure \ref{T21asymp} are the -- wrong -- predictions obtained by a theoretical analysis that would neglect the radiative modes. In such a case we have $|R|^2+|T|^2=1$ and \cite[Section 7.1.5]{book}:
\begin{equation}
    \label{eq:meanR2lambda0}
\mathbb{E}[|T|^2] \mid_{\Lambda=0} = \exp\Big(-\frac{\Ld}{\Ld_{\rm loc}}\Big) \int_0^\infty \exp\Big(-s^2 \frac{\Ld}{\Ld_{\rm loc}}\Big) \frac{2\pi s \sinh(\pi s)}{\cosh^2(\pi s) } ds .
\end{equation}
As seen in Figures \ref{T21asymp}(a-b) both the reflectance and transmittance are significantly over-estimated by this wrong approach.
It is, therefore, of utmost importance to take into account the coupling between surface and radiative modes, which can result in significant amount of radiative losses.

\subsection{Radiative loss: dependence on surface structure}
\begin{figure}[htb!]
\centering
(a)\includegraphics[width=.45\textwidth]{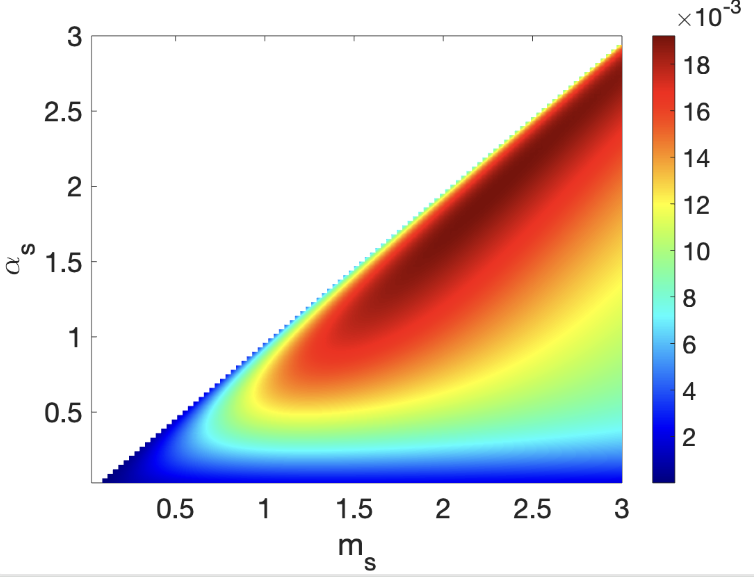}(b)\includegraphics[width=.45\textwidth]{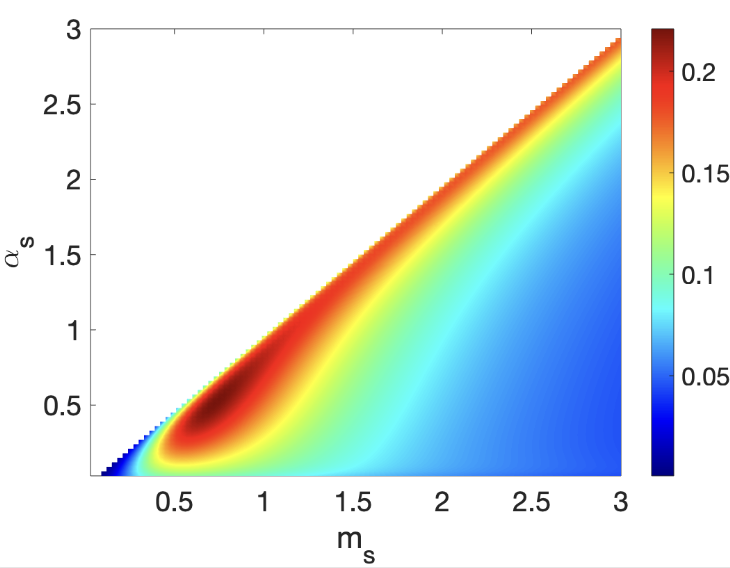}
\caption{
Mean radiative loss $\mathbb{E}[{\Dsf}]=1-\mathbb{E}[|R|^2+|T|^2]$ at frequency
(a) ${\omega}=0.5{\omega}_{\max}$ and (b) ${\omega}=0.9{\omega}_{\max}$, as a function of surface structure parameters $m_s$ and $\alpha_s$.
For these plots, $\sigma=0.05, \Ld=40$ and we use (\ref{eq:probaR2}) and (\ref{eq:probaT2}).
}
\label{fig:radloss1}
\end{figure}

\begin{figure}[htb!]
\centering
(a)\includegraphics[width=.45\textwidth]{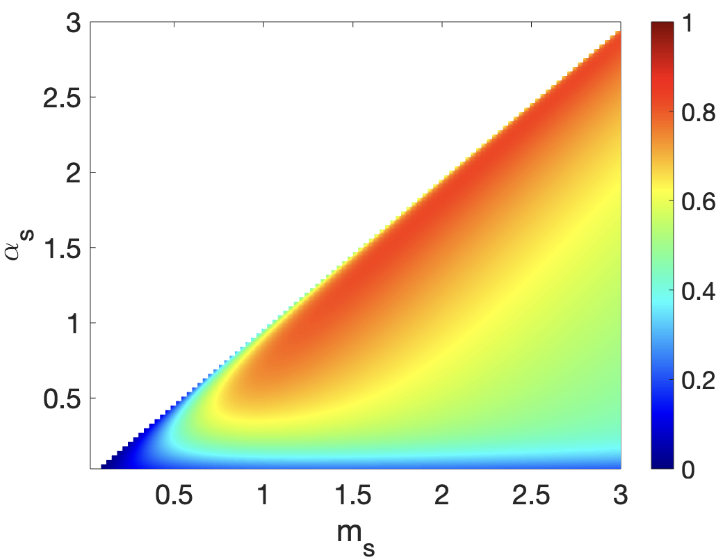}(b)\includegraphics[width=.45\textwidth]{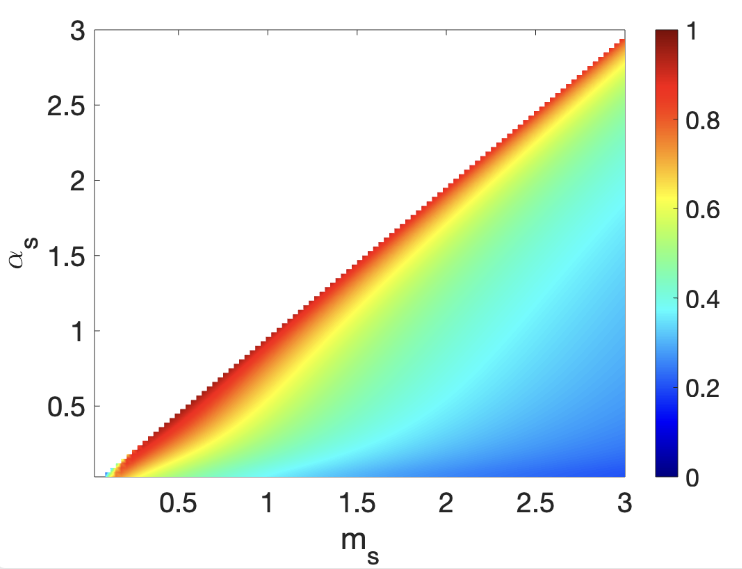}
\caption{
Mean radiative loss $\mathbb{E}[{\Dsf}]=1-\mathbb{E}[|R|^2+|T|^2]$ at frequency
(a) ${\omega}=0.5{\omega}_{\max}$ and (b) ${\omega}=0.9{\omega}_{\max}$, as a function of surface structure parameters $m_s$ and $\alpha_s$.
For these plots, $\sigma=0.05, \Ld=5000$ and we use (\ref{eq:probaR2}) and (\ref{eq:probaT2}).
}
\label{fig:radloss2}
\end{figure}

\begin{figure}[htb!]
\centering
(a)\includegraphics[width=.45\textwidth]{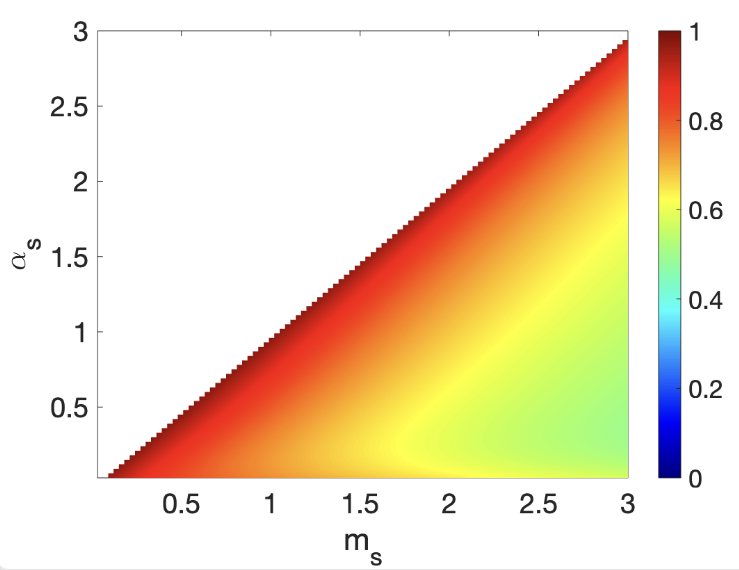}(b)\includegraphics[width=.45\textwidth]{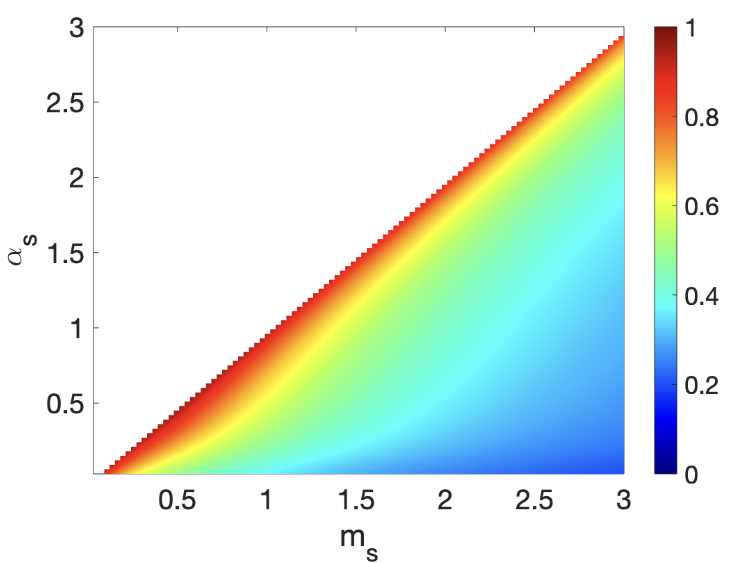}
\caption{
Mean radiative loss $\mathbb{E}[{\Dsf}]=1-\mathbb{E}[|R|^2+|T|^2]$ at frequency
(a) ${\omega}=0.5{\omega}_{\max}$ and (b) ${\omega}=0.9{\omega}_{\max}$, as a function of surface structure parameters $m_s$ and $\alpha_s$.
For these plots, $\Ld\to\infty$ and we use (\ref{ploteq1}).\\
}
\label{fig:radloss3}
\end{figure}

In Figures \ref{fig:radloss1}-\ref{fig:radloss3} we plot the mean radiative loss 
evaluated at ${\omega} =0.5{\omega}_{\max}$ (left) and ${\omega}=0.9 {\omega}_{\max}$ (right)
as functions of the surface structure parameters $m_s$ and $\alpha_s$ (recall that ${\omega}_{\max}$ is given by \eqref{defwmax}).
We can observe that the radiative loss becomes close to one in the strongly scattering regime and when $m_s$ and $\alpha_s$ are close to each other.
Indeed, when $m_s$ and $\alpha_s$ are close to each other,
the discrete eigenvalue of the operator \eqref{defLoper} is close to the continuous spectrum, or equivalently, the wavenumber of the surface mode is close to the wavenumber interval of the radiative modes. Under such circumstances, the coupling between the surface mode and the radiative modes induced by the mass perturbations is strong, which explains the value close to one of the radiative loss.

\section{Proof (multiscale analysis)}
\label{secmainproof}
\subsection{Modal expansion of the solution of the perturbed equation of motion}

We introduce the generalized forward-going (right-going) and backward-going (left-going) mode amplitudes,
\begin{equation}
\label{eq:amplitudes}
\big\{a_{\gamma_0}({\mathtt{x}}), \, b_{\gamma_0}({\mathtt{x}}) \big\} ~~ \mbox{and} ~~ \big\{a_{\gamma}({\mathtt{x}}), \, b_{\gamma}({\mathtt{x}}), ~ \gamma \in ({\omega}^2-4,{\omega}^2) \big\},
\end{equation}
which are defined by
\begin{align}
a_{\gamma}({\mathtt{x}}) =& \frac{e^{-i k(\gamma) {\mathtt{x}}}}{2i\sqrt{\sin k(\gamma)}} \Big( \check{\su}_{\gamma} ({\mathtt{x}}+1) -e^{-ik(\gamma) } \check{\su}_{\gamma} ({\mathtt{x}}) \Big), \\
b_{\gamma}({\mathtt{x}}) =& \frac{e^{i k(\gamma) {\mathtt{x}}}}{2i\sqrt{\sin k(\gamma)}} \Big( - \check{\su}_{\gamma} ({\mathtt{x}}+1) +e^{ik(\gamma) } \check{\su}_{\gamma} ({\mathtt{x}}) \Big), 
\end{align}
and which are such that 
\begin{align}
&
\frac{1}{\sqrt{\sin k(\gamma)}}\big( 
 {a}_{\gamma} ({\mathtt{x}}) e^{ik(\gamma) {\mathtt{x}}} +{b}_{\gamma}({\mathtt{x}}) e^{- ik(\gamma) {\mathtt{x}}} \big) =\check{\su}_{\gamma}({\mathtt{x}}), \\
&\big( {a}_{\gamma} ({\mathtt{x}}) -{a}_{\gamma}({\mathtt{x}}-1) \big) e^{ik(\gamma){\mathtt{x}}} 
+\big( {b}_{\gamma} ({\mathtt{x}}) -{b}_{\gamma}({\mathtt{x}}-1) \big) e^{-ik(\gamma){\mathtt{x}}} 
=0,
\end{align}
and 
\begin{align}
\nonumber
\check{\su}_\gamma({\mathtt{x}}+1)+\check{\su}_\gamma({\mathtt{x}}-1)+(\gamma-2) \check{\su}_\gamma({\mathtt{x}}) 
&
=
- 2i \sqrt{\sin k(\gamma)} e^{ik(\gamma){\mathtt{x}}} \big( a_\gamma({\mathtt{x}}-1)-a_\gamma({\mathtt{x}})\big)\\
&=2i \sqrt{\sin k(\gamma)} e^{-ik(\gamma){\mathtt{x}}} \big( b_\gamma({\mathtt{x}}-1)-b_\gamma({\mathtt{x}})\big).
\label{eq:radFB}
\end{align}
From (\ref{eq:cma2}) we obtain the coupled system of random difference equations 
satisfied by the mode amplitudes in \eqref{eq:amplitudes} for $\gamma \in ({\omega}^2-4,{\omega}^2)\cup \{\gamma_0\}$,
\begin{align}
\nonumber
& a_{\gamma}({\mathtt{x}}-1)-a_{\gamma}({\mathtt{x}})\\
\nonumber
& = - \frac{i m_{{s}}{\omega}^2\sqrt{\rho(\gamma)\rho(\gamma_0)}}{2 \sqrt{ \sin k(\gamma)\sin k(\gamma_0)}}
 {{\upmu}}_{{\mathtt{x}}} \Big[ {a}_{\gamma_0}({\mathtt{x}}) e^{i( k(\gamma_0) -k(\gamma)){\mathtt{x}}}
+ {b}_{\gamma_0}({\mathtt{x}}) e^{i( - k(\gamma_0) -k(\gamma)){\mathtt{x}}}
\Big]\\
\nonumber
&\quad - \frac{i m_{{s}}{\omega}^2 \sqrt{\rho(\gamma)}}{2 \sqrt{\sin k(\gamma)}} {{\upmu}}_{{\mathtt{x}}}
\int_{0}^{{\omega}^2} \frac{\sqrt{\rho(\gammap)}}{\sqrt{\sin k(\gammap)}} \Big[ {a}_{\gammap}({\mathtt{x}}) e^{i (k(\gammap)-k(\gamma)){\mathtt{x}}} +
 {b}_{\gammap}({\mathtt{x}}) e^{i (-k(\gammap)-k(\gamma)){\mathtt{x}}}
\Big]d\gammap\\
&\quad - 
 \frac{i m_{{s}}{\omega}^2 \sqrt{\rho(\gamma)}}{2\sqrt{\sin k(\gamma)}} {{\upmu}}_{{\mathtt{x}}}
 \int_{{\omega}^2-4}^0 \sqrt{\rho(\gammap)} \check{\su}_{\gammap}({\mathtt{x}}) e^{-i k(\gamma){\mathtt{x}}}
d\gammap,
\label{eq:evol1a}
\end{align}
\begin{align}
\nonumber
& b_{\gamma}({\mathtt{x}}-1)-b_{\gamma}({\mathtt{x}})\\
\nonumber
&= \frac{i m_{{s}} {\omega}^2\sqrt{\rho(\gamma)\rho(\gamma_0)}}{2 \sqrt{ \sin k(\gamma)\sin k(\gamma_0)}}
 {{\upmu}}_{{\mathtt{x}}} \Big[ {a}_{\gamma_0}({\mathtt{x}}) e^{i(k(\gamma_0)+k(\gamma)){\mathtt{x}}}
+ {b}_{\gamma_0}({\mathtt{x}}) 
 e^{i(-k(\gamma_0)+k(\gamma)){\mathtt{x}}} \Big]\\
\nonumber
&\quad + \frac{i m_{{s}} {\omega}^2 \sqrt{\rho(\gamma)}}{2 \sqrt{\sin k(\gamma)}} {{\upmu}}_{{\mathtt{x}}}
\int_{0}^{{\omega}^2} \frac{\sqrt{\rho(\gammap)}}{\sqrt{\sin k(\gammap)}} \Big[ {a}_{\gammap}({\mathtt{x}}) e^{i (k(\gammap)+k(\gamma))x}
+ {b}_{\gammap}({\mathtt{x}}) e^{i (-k(\gammap)+k(\gamma)){\mathtt{x}}}
\Big]d\gammap\\
&\quad + 
 \frac{i m_{{s}} {\omega}^2 \sqrt{\rho(\gamma)}}{2\sqrt{\sin k(\gamma)}} {{\upmu}}_{{\mathtt{x}}}
 \int_{{\omega}^2-4}^0 \sqrt{\rho(\gammap)} \check{\su}_{\gammap}({\mathtt{x}}) e^{i k(\gamma){\mathtt{x}}}
d\gammap .
\label{eq:evol1b}
\end{align}
This system is supplemented with boundary conditions corresponding to a unit-amplitude forward-going surface mode incoming from the left region ${\mathtt{x}}\leq 0$ and radiation condition into the right region ${\mathtt{x}} \geq \Ld+1$: 
\begin{equation}
\label{eq:bcabgamma0}
b_{\gamma}(\Ld+1) =0,\qquad a_{\gamma} (0)= {\bf 1}_{\gamma_0}(\gamma). 
\end{equation}

\subsubsection{Role of the evanescent modes}
The coupling with the evanescent modes
 can be captured by substituting the expression of the evanescent modes into the last terms of eqs. (\ref{eq:evol1a}) and (\ref{eq:evol1b}): for $\gamma \in ({\omega}^2-4,0)$, 
\[
\check{\su}_\gamma({\mathtt{x}}) = - \sum_{{\mathtt{y}}} {{\upmu}}_{{\mathtt{y}}} m_{{s}} {\omega}^2 \sqrt{\rho(\gamma) } \Big[ \sqrt{\rho(\gamma_0)} \check{\su}_{\gamma_0}({\mathtt{y}}) 
+ 
\int_{{\omega}^2-4}^{{\omega}^2} \sqrt{\rho(\gammap)}
\check{\su}_{\gammap}({\mathtt{y}}) d\gammap \Big] G_\gamma({\mathtt{x}}-{\mathtt{y}}),
\]
where $G_\gamma({\mathtt{x}})$ is the radiating solution of
\[
G_\gamma({\mathtt{x}}+1)+G_\gamma({\mathtt{x}}-1)+(\gamma-2) G_\gamma({\mathtt{x}})={\bf 1}_{0}({\mathtt{x}}),
\]
that is to say,
\begin{align}
\nonumber
G_\gamma({\mathtt{x}}) &= g_\gamma {\bf 1}_0 ({\mathtt{x}})
+
h_\gamma
e^{- k(\gamma) |{\mathtt{x}}|} (1- {\bf 1}_0({\mathtt{x}})),\\
g_\gamma &= \frac{1+e^{k(\gamma)} (\gamma-2)} {\gamma-2+(\gamma^2-4\gamma+2) e^{k(\gamma)} }, \\
\nonumber
h_\gamma &= - \frac{ e^{2k(\gamma)}} {\gamma-2+(\gamma^2-4\gamma+2) e^{k(\gamma)} },
\end{align}
for $\gamma \in ({\omega}^2-4,0)$.
The coupling with evanescent modes gives rise to an effective dispersion (i.e. a frequency-dependent phase modulation) as we will see below.

\subsubsection{Diffusion approximation}
It is not possible to apply directly diffusion approximation theory to the system (\ref{eq:evol1a}-\ref{eq:evol1b}) with boundary conditions (because the solution is not adapted to the filtration of the driving process ${\upmu}$, in the sense that the solution depends on $({\upmu}_{\mathtt{x}})_{ {\mathtt{x}} \in [1,\Ld]}$).
The strategy is to apply  diffusion approximation theory to the system (\ref{eq:evol1a}-\ref{eq:evol1b}) with initial conditions and to use the linearity of the system to deduce the behavior of the solution of the system with boundary conditions.

We first study the system (\ref{eq:evol1a}-\ref{eq:evol1b}) supplemented with initial conditions 
at ${\mathtt{x}}=0$ instead of (\ref{eq:bcabgamma0}), and we denote such a solution by $(a^{(i)}_{\gamma}({\mathtt{x}}),b^{(i)}_{\gamma}({\mathtt{x}}))$.
We apply the diffusion approximation theory set forth in \cite{book},
as in \cite[Appendix B]{Hoop}.
If $\Ld=[L/\sigma^2]$ and $\sigma^2 \to 0$,
then $(a^{(i)}_{\gamma_0}([x/\sigma^2]),b^{(i)}_{\gamma_0}([x/\sigma^2]))_{x\in [0,L]}$ converges to the solution $(\mathfrak{a}^{(i)}_{\gamma_0}(x),\mathfrak{b}^{(i)}_{\gamma_0}(x))_{x\in [0,L]}$ of the diffusion equation
\begin{align}
\nonumber
d
\begin{pmatrix}
\mathfrak{a}^{(i)}_{\gamma_0}(x)\\
\mathfrak{b}^{(i)}_{\gamma_0}(x)
\end{pmatrix}
=&
\frac{1}{\sqrt{L_{\rm loc}}}
\Big\{ \begin{pmatrix} i &0\\0 &-i \end{pmatrix} \begin{pmatrix}
\mathfrak{a}^{(i)}_{\gamma_0}(x)\\
\mathfrak{b}^{(i)}_{\gamma_0}(x)
\end{pmatrix}
dW_0(x) 
-\frac{1}{\sqrt{2}} 
\begin{pmatrix} 0&1\\1 &0\end{pmatrix} \begin{pmatrix}
\mathfrak{a}^{(i)}_{\gamma_0}(x)\\
\mathfrak{b}^{(i)}_{\gamma_0}(x)
\end{pmatrix}
dW_1(x)\\
&
+
\frac{1}{\sqrt{2}} 
\begin{pmatrix} 0 &-i\\i &0\end{pmatrix} \begin{pmatrix}
\mathfrak{a}^{(i)}_{\gamma_0}(x)\\
\mathfrak{b}^{(i)}_{\gamma_0}(x)
\end{pmatrix}
dW_2(x) \Big\}
+\frac{ \lambda+i\kappa}{2} 
\begin{pmatrix} -1&0\\0 &1\end{pmatrix} \begin{pmatrix}
\mathfrak{a}^{(i)}_{\gamma_0}(x)\\
\mathfrak{b}^{(i)}_{\gamma_0}(x)
\end{pmatrix}
dx
\label{eq:evoleabgamma0}
,
\end{align}
with the prescribed initial conditions and with
\begin{align}
\frac{1}{L_{\rm loc}}&=\frac{1}{\sigma^2 \Ld_{\rm loc}} = \frac{m_{{s}}^2 {\omega}^4 \rho(\gamma_0)^2}{4\sin^2(k(\gamma_0))},\\
\kappa &= \frac{m_{{s}}^2 {\omega}^4 \rho(\gamma_0)}{\sin k(\gamma_0)}
\int_{{\omega}^2-4}^0 \rho(\gammap) g_{\gammap} d\gammap,\\
\lambda &= \frac{\Lambda}{\sigma^2} = \frac{m_{{s}}^2 {\omega}^4 \rho(\gamma_0)}{\sin k(\gamma_0)}
\int_{0}^{{\omega}^2} \rho(\gammap)g_{\gammap} d\gammap.
\end{align}
Here $W_0$, $W_1$, $W_2$ are independent standard Brownian motions and $g_\gamma$ is defined by 
\begin{align}
\nonumber
g_\gamma &= i \frac{1+e^{-i k(\gamma)} (\gamma-2)} {\gamma-2+(\gamma^2-4\gamma+2) e^{-ik(\gamma)} } \\
&= \frac{1}{\sqrt{(4-\gamma)\gamma}} = \frac{1}{2\sin k(\gamma)}
, \quad \mbox{ for $\gamma \in (0,{\omega}^2)$,}
\end{align}
which is real, positive valued.
We can see that the coupling with the evanescent modes induces an effective dispersion term proportional to $\kappa$ and the coupling with the radiative modes induces an effective diffusion term proportional to $\lambda$.
We also remark that $\sigma^2$ does not appear in the expression of $L_{\rm loc}$, $\kappa$ and $\lambda$, but remember that $\Ld=[L/\sigma^2]$ and $(\mathfrak{a}^{(i)}_{\gamma_0}(x),\mathfrak{b}^{(i)}_{\gamma_0}(x))$ is the limit of $(a^{(i)}_{\gamma_0}([x/\sigma^2]),b^{(i)}_{\gamma_0}([x/\sigma^2]))$ as $\sigma \to 0$ so that 
$(a^{(i)}_{\gamma_0}({\mathtt{x}}),b^{(i)}_{\gamma_0}({\mathtt{x}})) \simeq (\mathfrak{a}^{(i)}_{\gamma_0}(\sigma^2 {\mathtt{x}}),\mathfrak{b}^{(i)}_{\gamma_0}(\sigma^2 {\mathtt{x}} ))$ for ${\mathtt{x}} \in [1,\Ld]$, $\Ld=[L/\sigma^2]$.

\subsubsection{Propagator, reflection and transmission}

We denote the solution of (\ref{eq:evoleabgamma0}) with the initial condition $\mathfrak{a}_{\gamma_0}^{(1)}(x=0)=1$, $ \mathfrak{b}_{\gamma_0}^{(1)}(x=0)=0$ by $(\mathfrak{a}_{\gamma_0}^{(1)}(x),\mathfrak{b}_{\gamma_0}^{(1)}(x))$.
Also, we denote the solution of (\ref{eq:evoleabgamma0}) with the initial condition $\mathfrak{a}_{\gamma_0}^{(2)}(x=0)=0$, $ \mathfrak{b}_{\gamma_0}^{(2)}(x=0)=1$ by
$(\mathfrak{a}_{\gamma_0}^{(2)}(x),\mathfrak{b}_{\gamma_0}^{(2)}(x))$.

By linearity of the system (\ref{eq:evol1a}-\ref{eq:evol1b}), if $\Ld=[L/\sigma^2]$ and $\sigma^2 \to 0$, then the mode amplitudes $(a_{\gamma_0}([x/\sigma^2]),b_{\gamma_0}([x/\sigma^2]))_{x\in [0,L]}$ (with $(a_{\gamma},b_{\gamma})$ solution of (\ref{eq:evol1a}-\ref{eq:evol1b}) with the boundary conditions (\ref{eq:bcabgamma0})) converges to $(\mathfrak{a}_{\gamma_0}(x),\mathfrak{b}_{\gamma_0}(x))_{x\in [0,L]}$ that satisfies 
\begin{equation}
\begin{pmatrix}
\mathfrak{a}_{\gamma_0}(x=L) =T \\
\mathfrak{b}_{\gamma_0}(x=L)=0
\end{pmatrix}
= 
\begin{pmatrix}
\mathfrak{a}^{(1)}_{\gamma_0}(x=L)&\mathfrak{a}^{(2)}_{\gamma_0}(x=L)\\
\mathfrak{b}^{(1)}_{\gamma_0}(x=L)&\mathfrak{b}^{(2)}_{\gamma_0}(x=L)
\end{pmatrix}
\begin{pmatrix}
\mathfrak{a}_{\gamma_0}(x=0)=1\\
\mathfrak{b}_{\gamma_0}(x=0)=R 
\end{pmatrix},
\end{equation}
where $T$ and $R$ are the limits of the transmission and reflection coefficients of the surface mode for the randomly perturbed region corresponding to a unit-amplitude right-going surface wave coming from $-\infty$ (i.e., boundary conditions (\ref{eq:bcabgamma0})).

We can also define the transmission and reflection coefficients corresponding to a unit-amplitude left-going surface wave coming from $+\infty$ (i.e., boundary conditions $b_{\gamma}(\Ld+1) ={\bf 1}_{\gamma_0}(\gamma)$, $a_{\gamma} (0)= 0$):
\begin{equation}
\begin{pmatrix}
\mathfrak{a}_{\gamma_0}(x=L) ={\tldR} \\
\mathfrak{b}_{\gamma_0}(x=L)=1
\end{pmatrix}
= 
\begin{pmatrix}
\mathfrak{a}^{(1)}_{\gamma_0}(x=L)&\mathfrak{a}^{(2)}_{\gamma_0}(x=L)\\
\mathfrak{b}^{(1)}_{\gamma_0}(x=L)&\mathfrak{b}^{(2)}_{\gamma_0}(x=L)
\end{pmatrix}
\begin{pmatrix}
\mathfrak{a}_{\gamma_0}(x=0)=0\\
\mathfrak{b}_{\gamma_0}(x=0)={\tldT} 
\end{pmatrix} .
\label{eq:relierefpropa}
\end{equation}
The pairs $(R,T)$ and $({\tldR},{\tldT})$ have the same distribution because $({{\upmu}}_{{\mathtt{x}}})_{{\mathtt{x}} =1}^{\Ld}$ has the same distribution as $({{\upmu}}_{\Ld+1-{\mathtt{x}}})_{{\mathtt{x}} =1}^{\Ld}$.
From (\ref{eq:relierefpropa}) the reflection and transmission coefficients $({\tldR},{\tldT})$ are given in terms of $(\mathfrak{a}^{(2)}_{\gamma_0}(x=L), \mathfrak{b}^{(2)}_{\gamma_0}(x=L))$ as
\[
{\tldR}= \frac{\mathfrak{a}^{(2)}_{\gamma_0}(x=L)}{\mathfrak{b}^{(2)}_{\gamma_0}(x=L)},\qquad 
{\tldT}= \frac{1}{\mathfrak{b}^{(2)}_{\gamma_0}(x=L)} .
\]
We then find from (\ref{eq:evoleabgamma0}) and It\^o's formula that, 
as a function of $L$, the pair $({\tldR},{\tldT})$ is a diffusion process which is solution of the stochastic differential equation:
\begin{align*}
d{\tldR} =& \frac{1}{\sqrt{L_{\rm loc}}}
\Big(2 i {\tldR} dW_0(L) -\frac{1}{\sqrt{2}} (1-{\tldR}^2) dW_1(L) - \frac{i}{\sqrt{2}} (1+{\tldR}^2) dW_2(L) \Big) \\
& - \frac{3}{L_{\rm loc}} {\tldR}dL 
- (\lambda +i\kappa) {\tldR}dL,
\end{align*}
\begin{align*}
d{\tldT} =& \frac{1}{\sqrt{L_{\rm loc}}}
\Big( i {\tldT}dW_0(L) +\frac{1}{\sqrt{2}} {\tldR} {\tldT} dW_1(L) - \frac{i}{\sqrt{2}} {\tldR}{\tldT} dW_2(L) \Big)\\
&- \frac{1}{L_{\rm loc}} {\tldT}dL
- \frac{1}{2} (\lambda +i\kappa) {\tldT}dL,
\end{align*}
starting from ${\tldR}(L=0)=0$ and ${\tldT}(L=0)=1$.
The application of It\^o's formula then gives that the moments of the reflection coefficient $\tilde{\mathcal R}_p = {\mathbb E} [ |{\tldR}|^{2p}]$ satisfy the system
\begin{equation}
\partial_L \tilde{\mathcal R}_p = \frac{p^2}{L_{\rm loc}} \big( \tilde{\mathcal R}_{p+1}+\tilde{\mathcal R}_{p-1} -2 \tilde{\mathcal R}_p)-2 \lambda p \tilde{\mathcal R}_p,
\end{equation}
starting from $\tilde{\mathcal R}_p(L=0) = {\bf 1}_0(p)$. This gives the desired result stated in Proposition \ref{prop:mainasy}.

\section{Conclusion}
Surface wave propagation across randomly perturbed surfaces of lattices is analyzed in this article. The perturbed constitution of surface associated with discrete Gurtin-Murdoch model, in contrast to uniform bulk elasticity, is captured by uniform mass and elastic constant for nearest-neighbour interactions parallel to the surface with a finite patch of point mass inhomogeneities.
The exact expression for surface wave reflectance and transmittance is provided for every such finite patch of mass perturbation on the surface. The exact semi-analytical expressions, as a result of stochastic, multiscale analysis for an ensemble of random mass perturbations, independent and identically distributed with mean zero, form the main result of this article.  
In the exposition, the cases of weakly and strongly scattering regimes are  included, which permit closed form expression for the moments of the radiative loss $\Dsf$. It is found that the mean value of $\Dsf$ shows distinct types of dependence of its behaviour on the assumed structure parameters.
In particular, the mean value of $\Dsf$ in weakly scattering regime is found to be proportional to an effective parameter that depends on the continuous spectrum of the unperturbed system, while in the strongly scattering regime, it is found to depend on another effective parameter but not on the variance of the mass perturbations. 
The theoretical predictions are supported by illustrations of their excellent agreements with numerical simulations for several choices of surface structure parameters.

\begin{figure}[htb!]
\centering
\includegraphics[width=.7\textwidth]{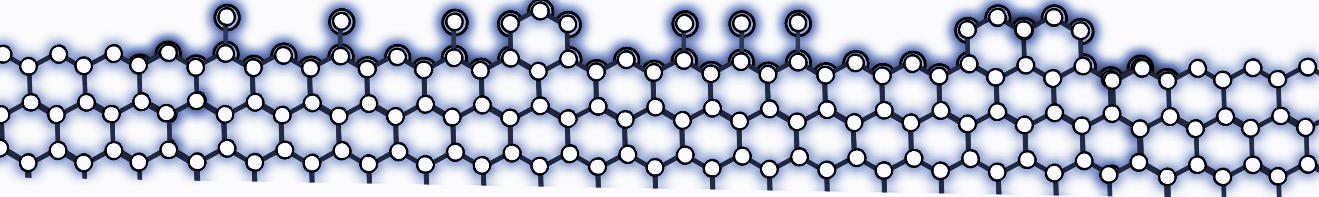}
\caption{Schematic of zigzag graphene edge with localized perturbations.}
\label{edgegraphenezigzag}
\end{figure}
Regarding the potential applications of the mathematical analysis of assumed prototype model, the considerations of this article also apply to electronic transport via edge states in Graphene-like structures \cite{Talirz,Merino,Wakabayashi,Baldwin} (see schematic in Fig. \ref{edgegraphenezigzag}). The tight-binding approximation in such situation, in the presence of surface perturbations, leads to equations similar to the ones analyzed in this article. A detailed and realistic model may not permit an exact theory as developed here but the qualitative nature of the results is anticipated to be the same; see, for example \cite{Blshexa,Blshexa2,Blsstep}, for some recent results involving the simplified tight binding models which utilize techniques similar to ones employed in the first part of the article.

\section*{Acknowledgments}
This work was started during the stay (in March 2023) of both authors at the Isaac Newton Institute (INI) for Mathematical Sciences, Cambridge. The authors would like to thank INI for support and hospitality during the programme -- `Mathematical theory and applications of multiple wave scattering' (MWS) where work on this paper was undertaken. This work was supported by EPSRC grant no EP/R014604/1. A part of the work of BLS, for the same visit to INI, was partially supported by a grant from the Simons Foundation.

\begin{appendices}

\section{Alternative derivation of the normalized eigenbasis}
\label{app:secGreen}
The following is an alternative derivation, using Green's function \cite{Stakgold}, to obtain the normalized eigenbasis described in Section \ref{seccomplete} for the semi-infinite (reduced) one dimensional lattice model. 

Based on \eqref{eqnhalf} and \eqref{bc1}, consider
\begin{equation}\begin{split}
(\Delta_1 +{\omega}^2)\psi({\mathtt{y}})=z\psi({\mathtt{y}}), 
\quad {\mathtt{y}}\in\mathbb{Z}^-,
\end{split}\end{equation}
\begin{equation}\begin{split}
\psi({\mathtt{y}}-1)-\psi({\mathtt{y}}) +{m}_{s}{\omega}^2\psi({\mathtt{y}})={\alpha}_{s} z\psi({\mathtt{y}})-1,
\quad {\mathtt{y}}=0.
\end{split}\end{equation}
Let
\begin{equation}\begin{split}
F_{s}:=(m_{s}{\omega}^2-\alpha_{s}\gamma-1).
\end{split}\end{equation}
Thus,
$\psi({\mathtt{y}})=\psi(0)\ell^{-{\mathtt{y}}},$
with $(\ell+F_{s}(z))\psi(0)=-1$,
where 
\begin{equation}\begin{split}
\ell+\ell^{-1}-2+{\omega}^2=z,
\text{ so that }\ell=\frac{1}{2}(z+2-{\omega}^2\pm\sqrt{(z+4-{\omega}^2)(z-{\omega}^2)}),
\end{split}\end{equation}
and sign chosen such that $|\ell|<1$ when ${\omega}$ is replaced by ${\omega}+i\varepsilon$ and $\varepsilon>0.$
With the point source at ${\mathtt{y}}=0$, in terms of the symbolic use of Green's function, let
$\psi({\mathtt{y}})=\mathtt{G}({\mathtt{y}},0;z).$
Thus,
\begin{equation}\begin{split}
\mathtt{G}({\mathtt{y}},0;z)=\frac{-1}{(\ell+F_{s}(z))}\ell^{-{\mathtt{y}}},\quad {\mathtt{y}}\le0.
\end{split}\end{equation}

Also, in terms of the eigenfunctions of $\mathcal{L}$ \cite{Stakgold},
\begin{align}
    \mathtt{G}({\mathtt{y}},0;z)={\rho({\gamma_{{\sw}}})}\frac{1}{z-\gamma_{\sw}}\psi_{\gamma_{\sw}}({\mathtt{y}})\overline{\psi}_{\gamma_{\sw}}(0)
+\int_{{\omega}^2-4}^{{\omega}^2} d\gamma {\rho({\gamma})}\frac{1}{z-\gamma}\psi_{\gamma}({\mathtt{y}})\overline{\psi}_{\gamma}(0),
\end{align}
i.e. as $\psi_{\gamma}(0)=1$,
\begin{equation}\begin{split}
\mathtt{G}({\mathtt{y}},0;z)={\rho({\gamma_{{\sw}}})}\frac{1}{z-\gamma_{\sw}}\psi_{\gamma_{\sw}}({\mathtt{y}})
+\int_{{\omega}^2-4}^{{\omega}^2} d\gamma {\rho({\gamma})}\frac{1}{z-\gamma}\psi_{\gamma}({\mathtt{y}}).
\end{split}\end{equation}
At $z=\gamma_{\sw}$, i.e. at the discrete eigenvalue, the residue 
\begin{align}
    Res_{z=\gamma_{\sw}} \mathtt{G}({\mathtt{y}},0;z)=\frac{-1}{(\ell(z)+F_{s}(z))'}\ell(z)^{-{\mathtt{y}}}|_{z=\gamma_{\sw}}, {\mathtt{y}}\le0,
\end{align}
can be simplified to
get
\begin{equation}\begin{split}
\rho({\gamma_{{\sw}}})=\frac{1}{\alpha_s+1/(\ell_{\sw}^{-2}-1)},\quad
\psi_{\gamma_{\sw}}({\mathtt{y}})=\ell_{\sw}^{-{\mathtt{y}}}.
\end{split}\end{equation}
For $\gamma\in({\omega}^2-4,{\omega}^2)$, i.e. on the continuous band,
$\mathtt{G}({\mathtt{y}},0;\gamma+i0)-\mathtt{G}({\mathtt{y}},0;\gamma-i0)
=-[\frac{1}{(\ell+F_{s}(z))}\ell^{-{\mathtt{y}}}]|^{\gamma+i0}_{\gamma-i0},$
which can be simplified to get
$\frac{i}{2\pi}(\mathtt{G}({\mathtt{y}},0;\gamma+i0)-\mathtt{G}({\mathtt{y}},0;\gamma-i0))=\frac{i}{2\pi}\frac{\ell_+-\ell^{-1}_+}{|\ell_++F_s|^2}\psi_{\gamma}({\mathtt{y}}),$
leading to
\begin{equation}\begin{split}
\rho({\gamma})=-\frac{i}{2\pi}\frac{\ell_+-\ell^{-1}_+}{|\ell_++F_s|^2}.
\end{split}\end{equation}

\section{Kolmogorov equations for jump Markov processes}
\label{app:jump}
A jump Markov process $(N_x)_{x\geq 0}$ with state space $\mathbb{N}$ is a stepwise constant Markov process which takes values in $\mathbb{N}$ \cite[Chapter 6]{book}. Its distribution is characterized by its generator ${\mathcal L}$ which is of the form 
\begin{equation}\begin{split}
{\mathcal L} f(n) = \lambda (n) \sum_{k\in \mathbb{N}} Q(n,k) \big(f(k)-f(n) \big) ,
\end{split}\end{equation}
for any test function $f$,
where for any integer $n$ $\lambda(n)>0$ and $k\mapsto Q(n,k)$ is a probability distribution on $\mathbb{N}$ (i.e., $Q(n,k)\geq 0$ and $\sum_{k\in \mathbb{N}}Q(n,k)=1$).

The generator makes it possible to compute any moment of the jump Markov process.
Indeed, if we denote $u(x,n) = \mathbb{E}[ f(N_x) |N_0=n]$, then $u$ satisfies the Kolmogorov equation 
\begin{equation}\begin{split}
\partial_x u = {\mathcal L} u, \quad x>0,
\end{split}\end{equation}
starting from $u(x=0,n)=f(n)$.

It is also possible to get a probabilistic representation for the solution of an equation of the form 
\begin{equation}\begin{split}
\partial_x u = {\mathcal L} u - V u, \quad x>0,
\end{split}\end{equation}
starting from $u(x=0,n)=f(n)$, where $V:\mathbb{N} \to [0,+\infty)$. Using Feynman–Kac representation formula, we have
\begin{equation}\begin{split}
u(x,n) = \mathbb{E}\Big[ f(N_x) \exp\Big(-\int_0^x V(N_y) dy \Big) \big| N_0=n\Big] .
\end{split}\end{equation}

The random dynamics of the jump Markov process $(N_x)_{x\geq 0}$ starting from $N_0=n$ is as follows:
\begin{itemize}
    
\item  sample a random variable $\tau_1$ with exponential distribution with parameter $\lambda(n)$, sample a random variable $Z_1$ with the distribution $Q(n,\cdot)$, and set $T_1=\tau_1$ and
\begin{equation}\begin{split}
N_x = n \mbox{ for } x \in [0,T_1).
\end{split}\end{equation}
\item  for $k\geq 2$, recursively, sample a random variable $\tau_k$ with exponential distribution with parameter $\lambda(Z_{k-1})$, sample a random variable $Z_k$ with the distribution $Q(Z_{k-1},\cdot)$, and set $T_k=T_{k-1} +\tau_k$ and
\begin{equation}\begin{split}
N_x = Z_{k-1} \mbox{ for } x \in [T_{k-1},T_k).
\end{split}\end{equation}
\end{itemize}

\end{appendices}

\bibliographystyle{siam}
\bibliography{./bibfile}

\end{document}